\documentclass{jfm}
\usepackage{graphicx}
\usepackage{epstopdf, epsfig,xcolor}

\shorttitle{Vortex Ring Interaction with Perforated plate}
\shortauthor{Jain, Rao and Basu}

\title{Interaction of vortex ring with perforated plate at different included angles}

\author{Siddhant Jain,
  Saini Jatin Rao,
 \and Saptarshi Basu\aff{}
\corresp{\email{sbasu@iisc.ac.in}}}
\affiliation{\aff{}Department of Mechanical Engineering, Indian Institute of Science,
Bengaluru, India
}
\begin{document}

\maketitle

\begin{abstract}

Experiments are performed to investigate the interaction of a vortex ring (Reynolds number based on circulation (\textit{Re$_{\Gamma}$} = 10500) with perforated surface (open area ratio, $\phi_1 = 0.24$ and $\phi_2=0.44$) with different included angles ($\theta$ = 60\textdegree-180\textdegree). The phenomenon is characterized using techniques like planer laser-induced fluorescence (PLIF) imaging and particle image velocimetry (PIV). Lagrangian analysis using finite-time Lyapunov exponents (FTLE) and $\Gamma_2$ vortex identification methods are utilized to understand flow physics. \textcolor{black}{We observe the development of mushroom-like structures at the holes, driven by the induced flow of the vortex ring. These formations, together with Kelvin-Helmholtz instability, introduce initial instability to the emerging jets. We discern a sequential emergence of the vortex ring in the form of jets at lower $\theta$ value that diminishes at higher values. Notably, a single vortex ring is split into two distinct vortex rings for $\theta$ $\leq$ 120\textdegree. On either side of the perforated plate, the sense of circulation after interaction does not show bias towards the sense of flow on the upstream region for some values of $\theta$. We further show the evolution of circulation by jets in the downstream region aligns with the proposed cumulative slug flow model using the centre line peak velocity of individual jets.}

\end{abstract}

\section{Introduction}
Vortex rings are formed as a result of an impulsive motion of fluid mass leading to the rolling up of shear layers at the corner of the pipe/hole used to generate it. They are fluidic structures that are of industrial as well as fundamental importance. In nature, they can be observed in smoke rings, jellyfish locomotion \citep{Dabiri2005}, wakes of flying birds \citep{Wolf2013-vy}, mushroom clouds during volcanic eruptions and explosion \citep{Lim1995}, micro-burst \citep{lundgren_yao_mansour_1992}, cardiac relaxation of human heart \citep{Arvidsson2016}, and more. Application-wise, they can be used for flow control purposes \citep{doi:10.2514/2.1323,YOU20081349}, heat transfer application of jets \citep{LIU19963695,Hadziabdic2008}, particle shaping technique \citep{An2016}, and cleaning purposes \citep{PhysRevFluids.8.044701} to name a few.

\textcolor{black}{It is important to study flow through perforated because of its high utility in the field of engineering. They find application in industrial processes like aggregation, sedimentation and filtration \citep{Narasimhamurthy_2022}. In addition to this, perforated plates are extensively used to model flow over wind turbines and wind farms \citep{Medici2005, Bossuyt2016, Steiros_Hultmark_2018} where the incoming flow field is rarely laminar. Since, it it difficult to get turbulent inlet conditions directly, vortical flows can act as an intermediate stage between laminar and turbulent flows \citep{Matsuzawa2023}. For flow control purposes, perforation on bluff bodies are used to reduce the drag by modifying the wake of the flow \citep{Castro_1971, Steiros_Hultmark_2018, Narasimhamurthy_2022}. Also, vortical flows through porous surfaces are important in context of human sneezing events, as was observed during COVID-19 pandemic time \citep{Bourouiba2014, Sharma_Sci_adv}.}

For a long time, physicists have been intrigued in understanding the fundamental behaviour of vortex rings and the interplay of their origin and decay with the surrounding. Works by \citet{Saffman1970, Maxworthy_1972, Maxworthy1977, Didden1979, Glezer1988, Glezer1990} include investigations on the fundamental aspects of vortex rings like the translation velocity, laminar and turbulent natures, their stability, and formation. More recent works by \citet{Gharib1998, Mohseni1998, Shusser2000, Gan2012, Krieg2013, Xiang2017} show the existence of a universal formation number, discuss formation time scales, drag forces, modeling of translation properties and energy of the vortex ring.

Of interest to the present work is how vortex rings interact with other bodies like vortex ring itself, solid or porous surfaces, and fluidic interfaces. \citet{Lim1992, Cheng2018} reported the collision of a vortex ring with another vortex ring giving rise to vortex re-connection and multiple small-sized vortex rings. \textcolor{black}{\citet{Walker1987} reported the development and separation of the boundary layer at the wall surface as a vortex ring approaches it in a normal direction. The separation caused the formation of a secondary and subsequently a tertiary ring. Numerical simulations carried out by \citet{Orlandi1990} for larger time periods showed multiple rebounds in case of vortex dipole interacting normally with plane wall with no-slip boundary conditions whereas under slip conditions, the rebounds were suppressed. \citet{Lim1989} investigated the interaction of a vortex ring with an inclined surface and reported the generation of bi-helical vortex line due to the variation of vortex stretching rate along the ring circumference. \citet{doi:10.1063/1.858650} highlights the importance of enstrophy over circulation in understanding various phases of a vortex ring as it approaches surfaces like plane walls or a free surface. A decrease in enstrophy during free-travelling stage due to viscous diffusion was observed, followed by a sudden increase due to vortex stretching and a decrease in total enstrophy during vortex rebound due to viscous dissipation. More recently, \citet{Couch2011} and \citet{New2016} used modern experimental facilities like planar laser-induced fluorescence (PLIF), digital particle image velocimetry (DPIV) and defocusing digital PIV (DDPIV) to quantify the events occurring when vortex ring comes in contact with inclined walls.
 Interactions of a vortex with round cylinders \citep{New2017} revealed the ejection of ringlets away from the cylinders due to strong interaction between primary and secondary vortex rings. \citet{ALLEN2007} quantified the motion of a free sphere on interaction with a vortex ring. The generation of crow-instability during the interaction of vortex ring with wavy wall was discussed by \citep{Morris_Williamson_2020}. \citet{PhysRevFluids.3.084701} experimented on the interaction of vortex ring with coaxial aperture of different sizes relative to the size of vortex ring. They proposed a semi-analytical model to predict the strength of the newly formed vortex ring for small aperture-to-ring radius cases. A study with non-permeable V shaped surface was conducted by \citet{New2020} where they explore the interaction for included angle (${\theta}$) = 30° - 120° at two different circulation based Reynolds number (\textit{Re$_{\Gamma}$}= $\Gamma/\nu$, $\Gamma$ is the circulation of the vortex ring and $\nu$ is the kinematic viscosity of the fluid): 2500 and 5000. A detailed discussion on the formation of secondary and tertiary vortex rings has been provided for orthogonal and valley planes (check figure 1(a) where \textit{xy}: valley plane and \textit{yz}: orthogonal plane).}
 
Works involving vortex ring interaction with porous surfaces have received relatively lesser attention. \citet{Adhikari_2009} conducted initial experiments to demonstrate the interaction of vortex ring (\textit{Re$_{\Gamma}$} =  384 to 2369) with porous surface (porosity = 62\% and 81\%) having constant mesh diameter (0.2mm). They showed similarities of such an interaction with the interaction of vortex ring with non-porous walls. \textcolor{black}{\citet{doi:10.1063/1.3695377} reported the formation of a coherent downstream vortex ring only in the case of fine mesh (woven stainless steel meshes of smaller wire diameter). In coarser mesh (larger wire diameter), chaotic downstream flow was observed because of small-scale vortical shedding from the cylindrical structures of the mesh.} \citet{naaktgeboren_krueger_lage_2012} proposed a model relating the kinetic energy dissipation and decrease in the hydrodynamic impulse of the vortex ring after interaction. \textcolor{black}{\citet{doi:10.1063/1.4897519} carried out a parametric numerical study and reported that reducing the porosity and increasing the thickness of porous surface had similar effects of reducing the vorticity transmission, contrary to an increase in the vorticity transmission by increasing the \textit{Re} of the vortex ring.} In another study by \citet{10.1115/1.4040215}, they argue that the formation of the downstream vortex ring is a function of vortex shedding at the mesh rather than screen porosity. \citep{xu_wang_feng_he_wang_2018} used the finite-time Lyapunov exponent (FTLE) and phase-averaged vorticity fields from the PIV data to reveal flow evolution in an interaction process. Further, they proposed a vorticity cancellation mechanism as the reason for the development of downstream vortex ring.
\citet{doi:10.1063/1.5100063} employed velocity triple decomposition to understand the interaction at different scales. They found that surface with largest hole resulted in minimum energy fluctuations calling it suitable for flow control.   The concept of vortical cleaning was recently introduced by \citet{PhysRevFluids.8.044701} where they utilized the kinetic energy of vortex ring to clean oil-impregnated porous surfaces of different porosity and shape. 

\textcolor{black}{From the literature survey conducted especially in the context of porous surfaces, the major gray regions can be identified: (i) The use of simpler porous surfaces/meshes/perforated plate/permeable surfaces with insignificant thickness (ii) The use of these surfaces with an included angle of 180\textdegree\ only i.e., flat surfaces. Furthermore, it has been reported that higher wire diameter induces more instability due to shedding type of phenomenon \citep{doi:10.1063/1.3695377, Cheng2010}. The authors believe this is somewhat analogous to the increased thickness of the perforated plates. However, the instability induced will now be a result of the shear layers developed inside the holes of larger thicknesses and not due to shedding. Motivated by this we use a perforated plate that hasn't been used before which provides significant resistance to the flow through it. Further, the utility of such a flow configuration can widely range for heat transfer to flow manipulation applications.}

\textcolor{black}{We consider a vortex ring with \textit{Re$_{\Gamma}$} = 10500 and two different open area ratios (open area/total area) ${\phi_{1}}$ = 0.24 and ${\phi_{2}}$ = 0.44 for the perforated plate that encompasses a broader range of spectra. The included angle, ${\theta}$ is varied from 60° - 180°, the latter being a plane perforated surface as shown in figure 1(d). We use  PLIF, PIV, FTLE fields, and $\Gamma_2$ vortex identification methods to bring out the flow physics involved during the interaction process.} \textcolor{black}{We further propose a cumulative slug flow model that uses the original slug model \citep{Didden1979} to relate the circulation generated in the downstream region with the time integral of the centre line velocity of each jets.}

The paper is organised as follows: Section 2 contains details of the Experimental setup and methodologies, section 3 discusses qualitative and quantitative results from the interaction process for both the perforated surfaces under subsequent subheadings, and section 4 presents a summary and conclusions of the present work.

\section{Experimental Setup}

\subsection{Flow chamber, vortex ring generation, and perforated surface}
All the experiments were conducted in a water-filled acrylic tank of size 450 mm x 450 mm x 600 mm having wall thickness of 8 mm with the free surface exposed to the atmosphere as shown in figure 1(a). The vortex ring of \textit{Re$_{\Gamma}$} = 10500 was generated using a solenoid valve connected to a pressurized water chamber. The solenoid valve, controlled by an Arduino board arrangement was opened for a specific time (60 ms) to ensure optimal mass ejection from an acrylic pipe of diameter (\textit{D}) 12 mm. The ejection pressure was set to 40 psi for all the cases. From the open end, around 25\% of the acrylic pipe was kept underwater to ensure no effects from the free surface. We use acrylic sheets of 3 mm thickness (\textit{h}) to make the perforated surface so as to facilitate optical accessibility required to conduct PIV and PLIF experiments. Two different perforated surfaces are considered in the present study having ${\phi}$ values of 0.24 (${\phi_1}$)and 0.44 (${\phi_2}$). This was achieved by keeping the hole diameter (${d_h}$) constant at 3 mm and changing the pitch (\textit{p}) as shown in figure 2(b). Two different perforated plates were attached at the centre aligning with the translational axis of the vortex ring to achieve the desired ${\theta}$ values. A frame with two side stands was designed as shown in figure 1(a) to hold the two ends of the perforated plate firmly inside the water. \textcolor{black}{The ratio of the area covered by the holes to the projected area of the vortex ring just before interaction was kept much larger than 1 (17 and 10.5 for $\phi_1$ and $\phi_2$, respectively).} The nomenclature used for each of the cases is in the form of ${\phi_{(1 or 2)}\theta^{(included angle)}}$ for example: ${\phi_{1}\theta^{60}}$ refers to the case with an open area ratio of 0.24 and included angle of 60\textdegree\ between the two surfaces. \textcolor{black}{For ease of understanding, the $\theta$ values are presented in their rounded-off form. Due to manufacturing inconsistencies, a maximum of $\pm$ 5\textdegree difference exists between the presented value and the actual value of $\theta$, with most cases having less than $\pm$ 3\textdegree.}

\begin{figure}
    \centering
    \centerline{\includegraphics[width=1\columnwidth]{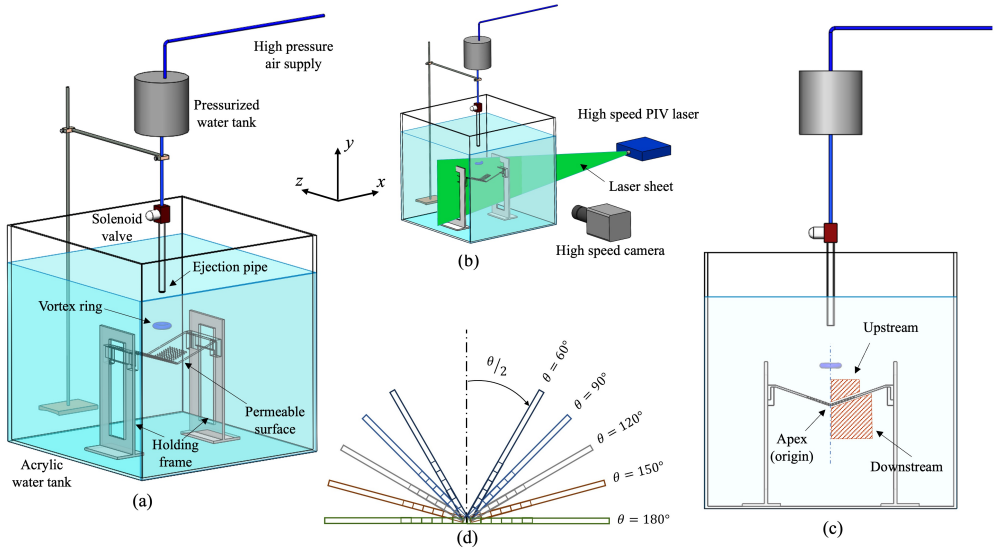}}
    \caption{(a) A 3-dimensional view of the experimental setup (b) Setup for PIV and PLIF measurements (c) A 2-dimensional view of the experimental setup. The hatched zone shows qualitatively the area considered for calculation of the circulation using the PIV data (d) Different configurations of the perforated surface used in the present study.}
    \label{figure:1}
\end{figure}

\subsection{PIV and PLIF}
The PIV measurements were done using a high-speed dual pulse Nd:YLF laser (Photonics Inc., pulse energy of 30 mJ, emission wavelength 527 nm). Both PIV and PLIF recordings were done at 1000 Hz. A typical setup for both PIV and PLIF experiments is shown in figure 1(b). For uniform illumination of the interaction zone, cylindrical lens beam  was expanded and converted into a planar sheet of thickness $\sim$ 1 mm. Neutrally buoyant borosilicate glass spheres (Sigma-Aldrich) of mean diameter ranging from 9-13 ${\mu}$m and having density of 1100 kgm$^{-3}$ were dispersed in water homogeneously and were illuminated by laser sheet. The seeder particles were also mixed with the water that was used to generate the vortex ring along with the quiescent fluid in the tank. The laser entered the acrylic tank from one side  (figure 1(b)) that was aligned precisely with the array of holes to acquire accurate flow field data coming out of the hole. \textcolor{black}{The images were recorded using Photron Mini UX high-speed camera at a pixel resolution of 1280 pixels x 1024 pixels with a field of view of $\sim$ 66 mm x 53 mm.} \textcolor{black}{Cross-correlation technique with decreasing multi-pass interrogation window size from 64 pixels x 64 pixels to 32 pixels x 32 pixels (wih 3 passes) was employed to process the displacement vectors. 50\% overlap was fixed for all the cases resulting in vectors at spacing of 0.84 mm in both x and y directions.}\textcolor{black}{A B-Spline bicubic reconstruction interpolation algorithm is performed for the final pass. A multi-pass median filter along with a denoising filter was used to eliminate the noise from the data. The uncertainty in velocity calculations was estimated using the Davis 8.4 (Lavision GmbH) to be below 12\%.} The captured PIV images were post-processed in Davis 8.4 software to obtain the velocity and vorticity fields that are subsequently utilized to construct FTLE fields and locate vortex core centres using $\Gamma_2$ method.

Additionally, for PLIF experiments, a small amount of rhodamine 6G dye was mixed with the fluid used to generate the vortex ring, and the emitted light was captured using the same camera mentioned above at 1000 Hz. A band-pass filter of 570 ± 10 nm (suitable for the emission range of fluorescent dye) was attached in front of the camera lens (LaVision) to block the scattered and stray light from entering the camera sensor.

\subsection{Lagrangian coherent structures (LCS) using FTLE}

In the present work, the finite-time Lyapunov exponents (FTLE) analysis is used to extract the LCS to study the interaction and mixing of the emerging jets from the perforated surface on interaction with the vortex ring. The FTLE fields are directly calculated using the velocity field obtained from the PIV data. FTLE is a scalar field that is used to measure at each point in space, the rate of separation of neighbouring particle trajectories initialized near that point \citep{Haller2001}. The method gained popularity in the last decade and is well documented \citep{10.1063/1.166479, SHADDEN2005271, doi:10.1146/annurev-fluid-010313-141322}. Under the action of a flow map ${F_{t_0}^{t_f}}$ the fluid particle x(${t_0}$) and x(${t_0}$)+ ${\delta}$(x(${t_0}$)) are defined in space and time. Using the local spatial gradients in the flow map i.e., ${\nabla}$${F_{t_0}^{t_f}}$(${x_0}$), the Cauchy-Green tensor is constructed \citep{10.1063/1.3579597}. The coefficient of expansion or the measure of strain is then defined as
\begin{equation}
\sigma(x_0,t_0) = {\lambda_{max}}([\nabla{F_{t_0}^{t_f}}({x_0})]^\otimes[\nabla{F_{t_0}^{t_f}}({x_0})])
\end{equation}
where $\otimes$ refers to the transpose of the matrix and $\lambda_{max}$ refers to the largest eigenvalue. The FTLE field is then defined as
\begin{equation}
FTLE(x_0,t_0) = \frac{1}{T}\log{\sigma({x_0,t_0})}
\end{equation}

\begin{figure}
    \centering
    \centerline{\includegraphics[width=1\columnwidth]{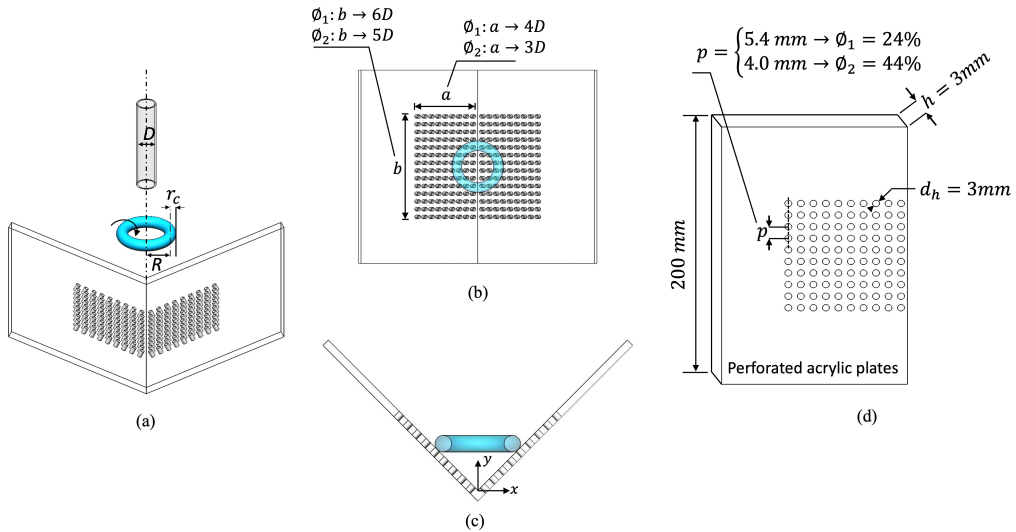}}
    \caption{ (not to scale) (a)A 3-dimensional schematic depicting various parameters of the vortex ring and ejection pipe (b) View from the frame of reference of the vortex ring before interacting (c) Side view of the vortex ring just before interacting (d) Dimensions of the two perforated surfaces used in the present study. (The number of holes and the lengths are just for representation and does not replicate the real scenario.}
    \label{figure:2}
\end{figure}

(For more details see \citep{SHADDEN2005271}). The maximum values of FTLE correspond to the LCS in the flow field. We calculate the FTLE in reverse time known as the backward FTLE that is associated with the convergence of the fluid particles. Higher values of FTLE mean a higher degree of convergence. The integration time \textit{T} can be chosen based on the extent the LCS needs to be captured \citep{shadden_katija_rosenfeld_marsden_dabiri_2007} however the trajectories should not leave the velocity field domain. We choose different values of \textit{T} for different frames cautiously keeping a check on the particle not getting accumulated at the perforated surface and not leaving the flow domain. Further, the value of \textit{T} should not be kept smaller than the time scales of the formation of the coherent structures. Finer spatial grids result in better and sharper fields \citep{Vetel2009,ESPA20129}. In the present work, the grid resolution was increased to obtain sharper LCS ridges. The FTLE ridges can be successfully utilized to identify boundaries of the coherent structure in both laminar or turbulent flows like oceanic flows \citep{https://doi.org/10.1002/2013GL058624}, large scale environmental flows \citep{Olascoaga2012}, mixing in turbulent flows \citep{PhysRevLett.98.144502}, laminar flow structures like periodic shedding of vortices behind bluff bodies \citep{10.1063/1.3483220}, vortex ring formation in heart \citep{ESPA20129} and more \citep{doi:10.1146/annurev-fluid-010313-141322}.

\subsection{${\Gamma_2}$ method for vortex core identification}
The use of vorticity fields to detect vortices can lead to erroneous interpretations \citep{Hussain1995}. Since, vorticity calculation involves velocity gradients, the vorticity fields can fail to differentiate between shear layers and rotating regions. In the present work, we use vorticity contours to calculate the circulation and ${\Gamma_2}$ method to educe the centres of rotating vortices. ${\Gamma_2}$ method, as proposed by \citet{Graftieaux_2001} is defined as

\begin{equation}
\Gamma_2(P) = \frac{1}{N}\sum_S \frac {({\vec{PQ}} \times ({\vec{U_Q} - \vec{U_m}}))\cdot \vec{z}}{\parallel \vec{PQ} \parallel\cdot \parallel \vec{U_Q}-\vec{U_m} \parallel}
\end{equation}

where the formulation is applied at each point P in the vector field obtained from the PIV data to find the normalized scalar $\Gamma_2$ values. $\vec{PQ}$ denotes the displacement vector from point P to Q inside area S, $\vec{U_m}$ refers to the mean velocity of the area S, $\vec{U_Q}$ refers to the velocity vector at point Q and $\vec{z}$ is the unit vector perpendicular to the plane. The threshold for the vortex centre is considered to be 0.75 (i.e., $>$ 2/$\pi$) since the structures formed in the downstream region after interaction are chaotic. The consideration of the local velocity vector $\vec{U_m}$ makes the method Galilean invariant unlike $\Gamma_1$ method \citep{Graftieaux_2001}.

\begin{table}
  \begin{center}
\def~{\hphantom{0}}
  \begin{tabular}{lcccccccc}
     \textit{P$_{in}$}\hspace{4pt} & \textit{r$_c$} \hspace{4pt}  & \textit{R} \hspace{4pt} & \textit{U$_c$} \hspace{4pt}& \textit{${\Gamma_{0}^+}$} \hspace{4pt} & \textit{${\Gamma_{0}^-}$} \hspace{4pt} & \textit{Re$_{\Gamma}$} \hspace{4pt} & \textit{Re$_{conv}$}\\[4pt]
    (psi) & (cm) & (cm) & (cm/s) & {($cm^{2}/s$)} & {($cm^{2}/s$)} & (\textit{${\Gamma_{0}/\nu}$}) & (\textit{${U_{c}R/\nu}$}) \vspace{6pt} \\
     
       40   & 0.4 $\pm$ 1\% & 0.9 $\pm$ 1.6\% & 30 $\pm$ 3\% & 105.5 $\pm$ 3\% & 103.7 $\pm$ 2.9\% & 10500 & 2460  \\

  \end{tabular}
   \caption{Different parameters of the vortex ring. The $\pm$ \% indicates the positive and negative standard deviation values of the corresponding mean values. Here, \textit{P$_{in}$} refers to the injection pressure at which the vortex ring is generated inside the water; \textit{r$_c$} refers to the radius of the core of the vortex ring (check figure 2(a)); \textit{R} is the distance between the vortex centre to the core known as the radius of the vortex ring (check figure 2(a)); \textit{U$_c$} is the convection velocity of the vortex core in y direction calculated by tracking the vortex core; \textcolor{black}{\textit{${\Gamma_{0}^+}$} and \textit{${\Gamma_{0}^-}$} are positive and negative circulation of the free vortex ring} i.e., the case without any perforated plate; \textit{Re$_{\Gamma}$} is Reynolds number based on circulation; \textit{Re$_{conv}$} is the Reynolds number based on convection velocity and ring radius.} 
 \label{tab:kd2}
  \end{center}
\end{table}

\subsection{Experimental conditions and measurements}
The experiments are conducted underwater at room temperature. For all the values of ${\phi}$ and ${\theta}$ the distance between the ejection pipe and the apex of the perforated surface was kept at 7\textit{D}. This distance is kept such that the vortex ring is fully developed during the start of the interaction. It is worth noting that in the present scenario, at lower ${\theta}$, the vortex ring will start interacting earlier compared to larger ${\theta}$ values and hence the apex is considered as the reference point from which the y axis passes. This axis also coincides with the translational axis of the vortex ring. The various specifics of the vortex ring are enumerated in Table 1 and have been depicted in Figure 2(a). The velocity and vorticity (${\omega}$) fields obtained from the PIV measurements are used to calculate the circulation (${\Gamma}$) by taking the area integral of the vorticity field. The formulation for circulation is given as

\begin{equation}
\Gamma =\int_C \omega dA
\end{equation}

where \textit{C} is the area of interest. For all the cases, a threshold of 10\% of the maximum vorticity was considered to calculate the circulation which was found to be sufficient to eliminate the background noise in the PIV data. For calculating the circulation, a masking region was defined as shown in Figure 1(c), encompassing the desired events. \textcolor{black}{The velocity magnitude of the free vortex ring that is used to normalize the velocity data (in figure 8) is calculated by taking the mean value of the velocity magnitude values along the line connecting the two cores.} \textcolor{black}{Further, for each of the data presented, the values have been averaged over 5 runs to maintain repeatability. Since, the phenomenon is sensitive to the alignment of the vortex ring with the apex of the perforated plate, special care was taken to check the symmetricity of the interaction for each run.}

\section{Results and Discussion}

\subsection{Characterization of the vortex ring}

Vortex ring having \textit{Re$_{\Gamma}$} = 11500 {(at 5\textit{D} from ejection pipe)} is considered in the present study to understand how ${\theta}$ and ${\phi}$ influences the interaction. Figure 3(a) depicts the variation of ${\Gamma^{+}_0}$ and ${\Gamma^{-}_0}$ (circulation generated by positive and negative vorticity respectively) of the free vortex ring (without perforated surface) with \textit{ T$^*$} (non-dimensionalized using the ratio \textit{${U_{c}/D}$}). This period corresponds to a physical distance of $\sim$ 3 cm (2.5\textit{D}) starting at 5\textit{D} from the ejection site. The constant value of circulation confirms the fully developed condition of the vortex ring before the interaction. Figure 3(b) shows the y-velocity profile of the vortex ring along the line joining the two cores as shown in the inset. The profile obtained corresponds to a typical velocity profile across the vortex ring \citep{Maxworthy1977}. From the profile, the symmetric nature of the vortex ring can be confirmed. Figure 3(c) depicts the velocity vector field of a propelling vortex ring superimposed over the vorticity contours. The centres of the two cores are identified using the $\Gamma_2$ method as described in section 2.4. It is clear that the peak vorticity and the centre do not coincide. The backward time FTLE field of the vortex ring is shown in figure 3(d). The green ridges corresponding to the higher FTLE values demarcate the vortex ring boundary across which material transport remains restricted.

\begin{figure}
    \centering
    \centerline{\includegraphics[width=1\columnwidth]{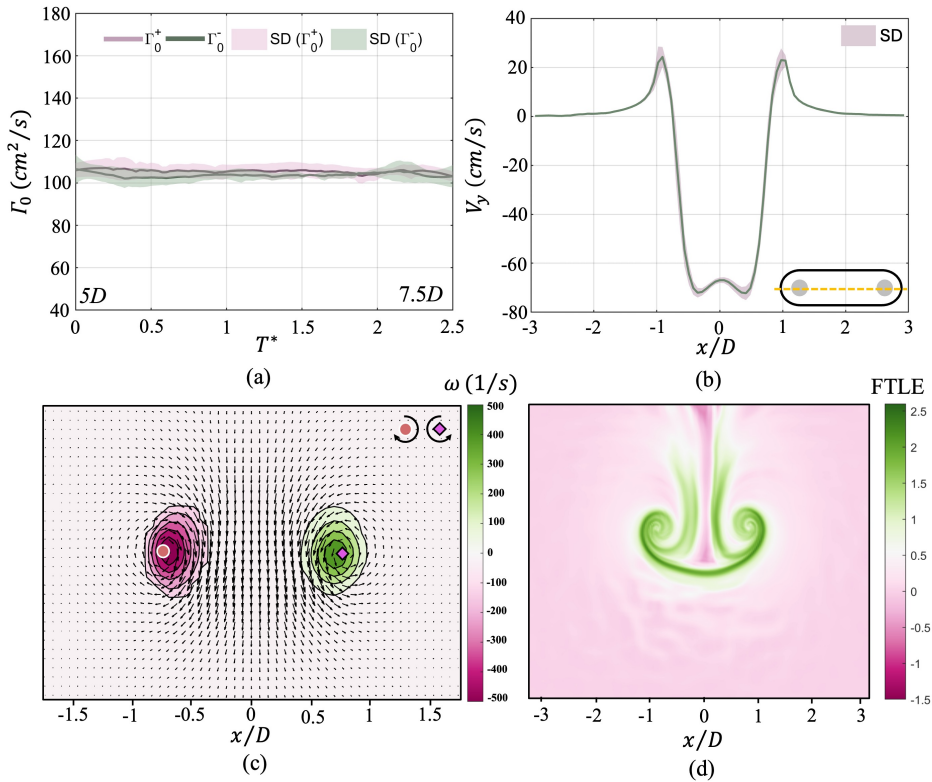}}
    \caption{(a) Variation in the positive and negative circulation of the free vortex with the non-dimensional time (${T^{*}=\textit{${U_{c}/D}$}}$). The distance covered during this period  starts from 5D till 7.5D from the ejection pipe (b) The y-velocity ($V_y$) profile of a free vortex at 5D distance from the ejection pipe along the line passing through the vortex core as shown in the inset (c) Vorticity contour plot of a free vortex ring overlaid over velocity vectors field. The pink and green dot represents the counter-clockwise (CCW) and clockwise (CW) vortex core detected using $\Gamma_2$ method (d) The backward FTLE field of a free vortex overlaid over the velocity vector field.}
    \label{figure:3}
\end{figure}

\subsection{Qualitative aspects of the interaction}

To develop an initial understanding of the interaction, we show in figure 4 and 5 the time-sequenced PLIF images at different non-dimensional times (\textit{T$^*$}= \textit{${tU_{c}/D}$}))  for all the ${\theta}$ values at ${\phi_1 =0.24}$ and ${\phi_2 =0.44}$. \textit{{T$^*$}}= 0 is considered just before the dyed vortex collides with the perforated wall. Due to hindrance from the perforated surface, the left side of the interaction process is not visible or partly visible hence, the right side is considered for discussion. An important aspect of increasing the ${\theta}$ values is that it delays the interaction phenomenon and shifts the interaction site further downward. This combined with the ${\phi}$ has a direct consequence on the number of holes involved during the interaction process. For ${\phi_1\theta^{60}}$ (Figure 4), the vortex ring on interaction starts emerging out as unstable jets from the $4^{th}$ hole counted from the apex. This number decreases by 1 till ${\phi_1\theta^{120}}$ after which it emerges from 2 holes. For $\phi_2$, the number of holes remain 5 for ${\phi_2\theta^{60}}$ that decreases by 1 till ${\phi_1\theta^{150}}$. This is because of the smaller pitch between holes for ${\phi_2}$ compared to ${\phi_1}$ (check figure 2). These observations will be quantitatively explained in further sections. For both ${\phi_1\theta^{180}}$ and ${\phi_2\theta^{180}}$ the vortex ring comes out of 5 holes however, shifting the position of the perforated surface for ${\phi_2\theta^{180}}$ results in involvement of 6 holes. For consistency, we ignore this case in the present study and consider the case when the translational axis of the vortex ring coincides with one of the hole centres for ${\phi_2\theta^{180}}$.

\begin{figure}
    \centering
    \centerline{\includegraphics[width=1\columnwidth]{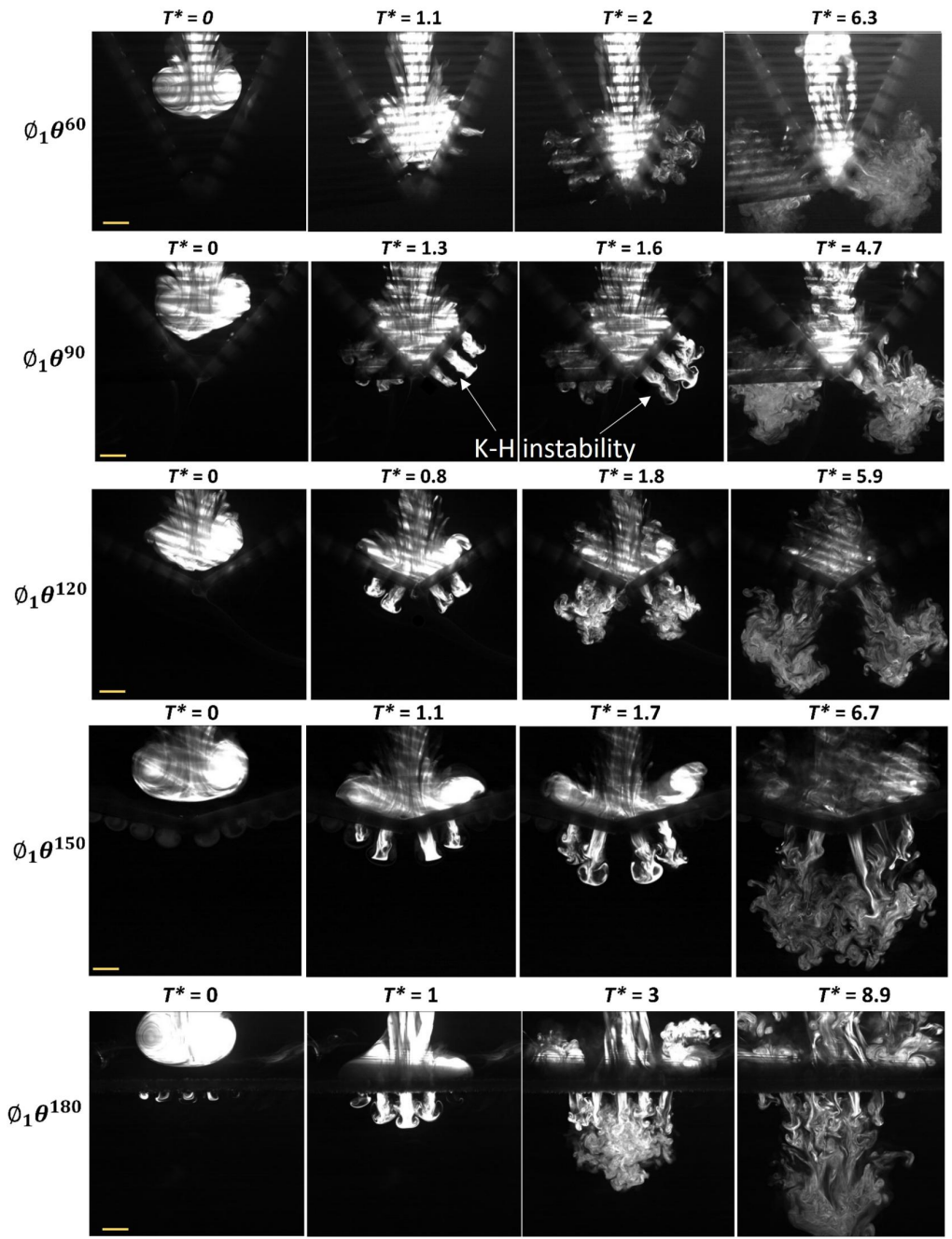}}
    \caption{The PLIF images for ${\phi_1}$ at ${\theta}$ = 60\textdegree\ - 180\textdegree\ shown at different ${T^{*}}$. The scale represents 5 mm.}
    \label{figure:4}
\end{figure}
\begin{figure}
    \centering
    \centerline{\includegraphics[width=1\columnwidth]{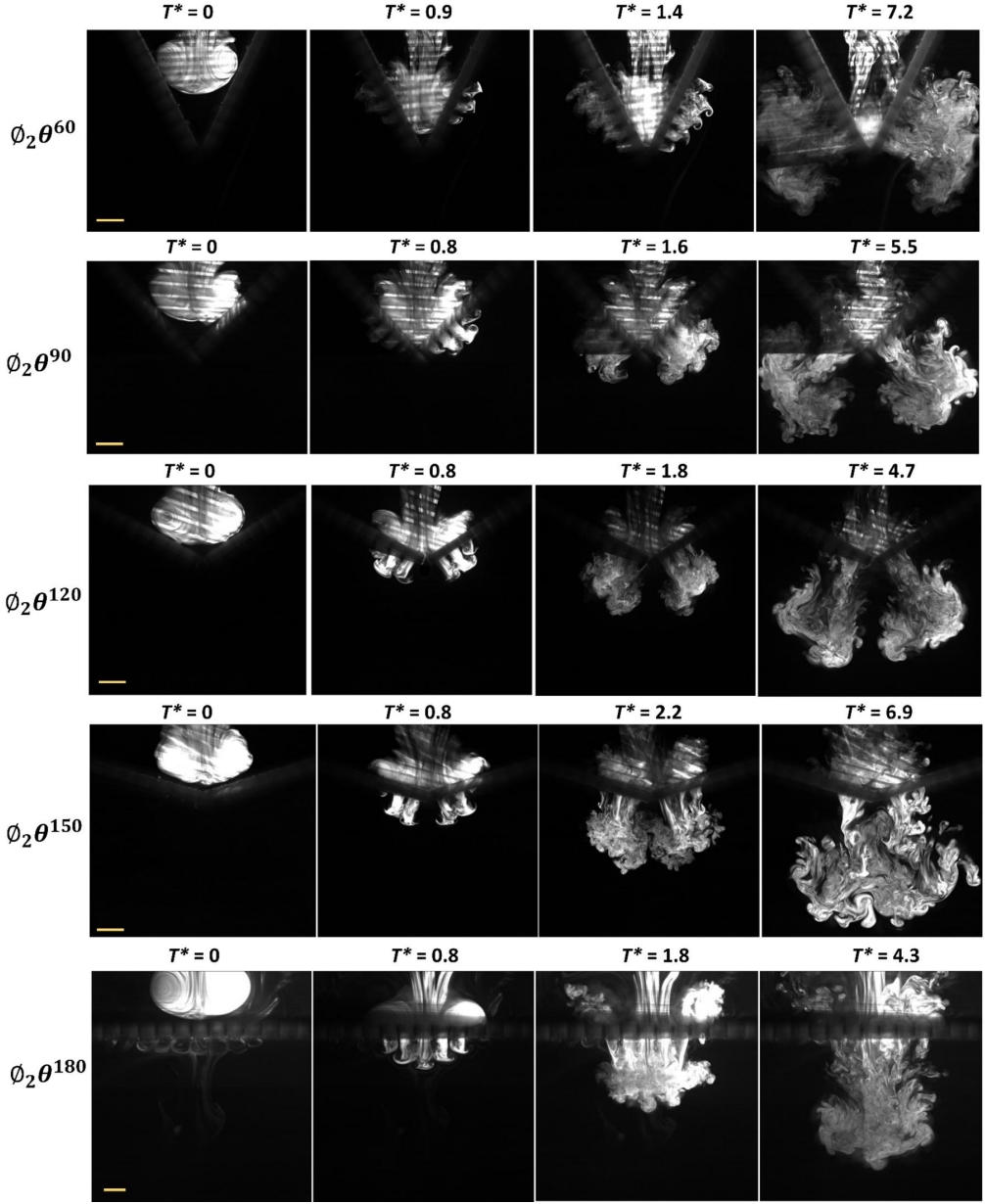}}
    \caption{{The PLIF images for ${\phi_2}$ at ${\theta}$ = 60\textdegree\ - 180\textdegree\ shown at different ${T^{*}}$. The scale represents 5 mm.}}
    \label{figure:5}
\end{figure}
\begin{figure}
    \centering
    \centerline{\includegraphics[width=1\columnwidth]{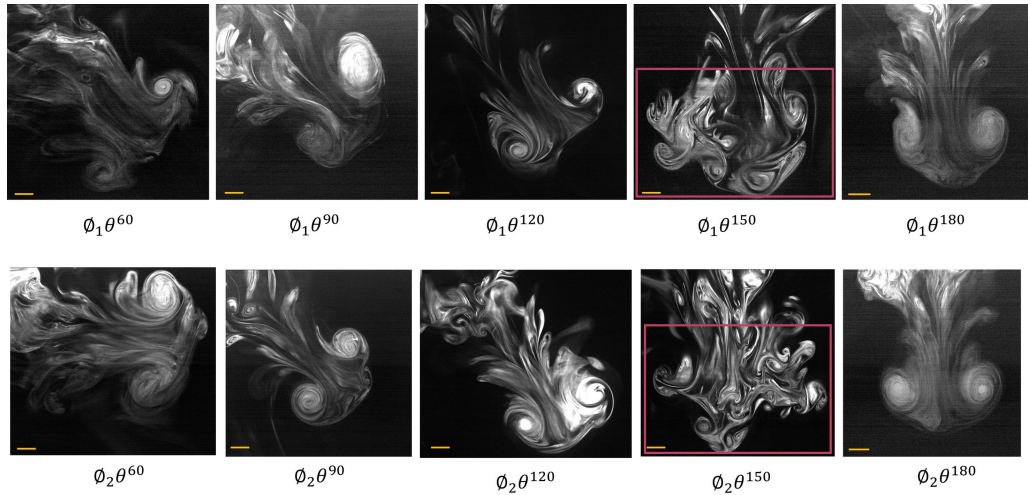}}
    \caption{{The reformation of the vortex ring in the far downstream region ($\sim$  6\textit{D} from the apex, except for $\theta$ = 150\textdegree). The red box in the case of $\theta$=150\textdegree\ depicts the suppression of the reformed vortex ring. The scale bars represent 5 mm.}}
    \label{figure:6}
\end{figure}

On interacting, the vortex ring breaks down into circular jets that develop instability over very small time scales. Due to the quiescent nature of fluid downstream of the surface, Kelvin-Helmholtz (K-H) instabilities grow at the shear layers of the jets that are clearly visible in the figures (4-5). For cases with lowest $\theta$ value (i.e., 60$^{\circ}$), the flow after interaction has a high tendency to curl in the same sense as the right-hand part of the vortex ring (as will be discussed in section 3.3.2, check figure 14). The emerging  jets start travelling in x direction with their heads rotating in an opposite sense. This can be clearly seen in ${\phi_2\theta^{60}}$ (Figure 5) where the head of the jets roll up to form series of circular patterns, a feature very commonly observed during the occurrence of K-H instabilities. Furthermore, for ${\phi_1}$ the out-coming jets maintain a distance with the adjacent jets for a long time before interacting as is visible in Figure 4 (${\phi_1\theta^{60}}$: \textit{{T$^*$}}=1.1, ${\phi_1\theta^{90}}$:\textit{{T$^*$}}=1.3, 1.6, ${\phi_1\theta^{120}}$:\textit{{T$^*$}} = 0.8, ${\phi_1\theta^{150}}$: \textit{{T$^*$}}=1.1, 1.7, ${\phi_1\theta^{180}}$: \textit{{T$^*$}}=1). Whereas in cases with ${\phi_2}$, these jets interact with their adjacent jets very early because of the proximity of the holes. These early interactions make the downstream flow highly chaotic in nature. Adding on to the instabilities, we observe an intriguing phenomenon that results in the mushroom-shaped features at the head of the jets that will be discussed in section 3.2.1.

The jets after ejecting out self interact resulting in a violent mixing of the fluid that has been observed in previous studies \citep{xu_wang_feng_he_wang_2018, PhysRevFluids.8.044701}. This interaction is characterized by 3-dimensional mixing and vorticity cancellation. The consequence of vorticity cancellation is the formation of two primary coherent structures in the downstream with low strength. With increasing ${\theta}$, the fluid from the two halves of the surface starts to approach each other and interacts for ${\theta}$=150\textdegree\ (figures 4 and 5, $\theta$=150\textdegree). The distance at which the two halves interact is not sufficient for each half to form the downstream vortex hence, we did not observe a collision of two downstream regenerated vortices. Due to this type of interaction, no coherent structure is observed for ${\phi_1\theta^{150}}$ cases. For all the other cases the regenerated downstream vortex is shown in Figure 6. 

It can be understood that on interaction, the vortex ring will expand in the z-direction as well as has been discussed by \citet{New2020} in detail. In their study, the walls had no penetration condition which forced the vortex ring to expand in the z direction. In the present study due to the change in the penetration condition at the wall, we see a rapid passage of vortex through the holes in a sequential manner as discussed above. An image (see supplementary figure S1) has been provided to illustrate the amount of dyed fluid ejecting of the holes present in the z direction from the apex for ${\phi_1\theta^{90}}$. The image shows that for lower open area ratio i.e., ${\phi_1}$, the amount of fluid coming out at different distances remains significantly less. \textcolor{black}{Qualitative images of the flow in z-direction are also shown by imaging the interaction process at the plane of symmetry which was visualised by placing the camera inclined at 45\textdegree\ angle.} Essentially, this must depend strongly on the ${\phi}$ value of the system and is expected to be much more lesser for ${\phi_2}$.

\subsubsection{Initial instability on jets}

\textcolor{black}{It is well known that a vortex ring induces fluid motion around itself. The velocity field around a vortex decreases gradually as we go away from its central region (the gradual decrease can be seen in figure 3(b).} Due to the induced flow and the presence of perforated surface of appreciable thickness, the fluid ahead of the vortex ring crosses the perforated surface much before the vortex ring does. As the induced flow passes through the holes, the shear resistance offered by the holes leads to the development of mushroom shape front. This can be seen in figure 7 where we capture this phenomenon for ${\theta}$ = 150\textdegree\ by pouring a thin layer of rhodamine dye over the perforated surface just before the vortex ring approaches it. This front is further caught up by the main flow from the vortex which can be seen from the markings in figures 7. The initial jets have low kinetic energy as the peak comes later (\textit{{T$^*$}} $>$ 1, check figure 8) due to penetration of the core region and hence this initial jet converges to the shape of this mushroom. This imparts the initial instability to the jet heads, forming a mushroom-shaped feature. In case of ${\phi_2}$, these mushrooms remain in close proximity tending to overlap at the edges thereby forming a continuous structure (figure 7(b)), unlike in ${\phi_1}$ where the boundaries of the mushrooms are distinctly detectable. \textcolor{black}{As the dominant part of the vortex ring (the core region that has higher vorticity) starts to penetrate, a velocity peak is obtained, and intense mixing between jets is observed along with K-H instability at the edges of the jets.} The initial instability discussed here adds to the rolling up of the jets at their heads. This type of mushroom formation is a direct consequence of placing a perforated plate with considerable thickness in the path of a vortex ring. \textcolor{black}{For further visualization of the velocity field around the vortex ring and the mushroom formation, check the supplementary figures S2, S3 and S4.} \textcolor{black}{The authors found no discussion of such events in the existing literature and speculate that this might be due to variations in the thickness and types of porous or perforated plates used in other studies. However, our research highlights the significance of these mushroom-shaped structures in understanding jet instability, raising the question of whether vortex rings should be treated as discrete systems in interaction studies.}

\subsection{Quantitative aspects of the interaction}

\subsubsection{Evolution of velocity and vorticity field near the surface}

As seen in the preceding sections, the vortex ring on colliding with the inclined perforated surface emerges out sequentially. To better understand the phenomenon we plot the non-dimensional velocity magnitude (${V_{mag}^*}$, normalised using the velocity magnitude of the free vortex) and the dimensional vorticity values about a line at a distance of $\sim$ 0.5\textit{D} from the perforated surface in the downstream region as shown in figure 8 and 9 respectively. This distance was considered to avoid the contamination of the data by the noisy signals generated near the wall. Also, beyond this distance, some of the peaks were lost in few cases. In the y-axis the non-dimensional length ($L^*=L/D$) with 0 being the apex point is plotted along with the non-dimensional time ($T^*=tU_{c}/D$) in the x-axis. $T^*=0$ is considered the frame just before the velocity vectors start emerging from the perforated surface in the PIV data. The primary purpose of these plots is to depict the sequential generation of jets from the perforated plate and to capture the near-wall events when the jets emerge out. For ${\phi_1\theta^{60}}$, we see a weak $V_{mag}^{*}$ peak followed by three dominant peaks occurring one after the other as time progresses (figure 8). Although for ${\phi_1\theta^{60}}$ the vortex interacts early with the wall, the initial impact remains weak and vortex rapidly gets sucked in towards the apex. The initial interaction does not involve the disintegration of the vortex core resulting in a feeble peak. Overall, we see three major peaks for this case that slowly convects out with time. Similarly, for ${\phi_2\theta^{60}}$, the interaction starts even earlier which is expected and 5 distributed peaks are obtained in a successive manner (figure 8). The phenomenon seen in the PLIF images (figure 4 and 5) for ${\theta}$=60\textdegree\ has been well captured by the PIV data as can be seen from the vorticity values plotted in figure 9. For ${\phi_1\theta^{60}}$ we can see the appearance of alternate sense of vorticity patches i.e., positive vorticity (red contours, CCW) sandwiched between negative vorticity zones starting from the farthest hole with weak negative vorticity zone. Whereas for ${\phi_2\theta^{60}}$, we can see a continuous stretch of CW vortices appearing initially followed by a slightly shifted stretch of CCW vortices. The vorticity zone near the apex maintains its vorticity for the longest period due to the decay of the vortex near the apex.

\begin{figure}
    \centering
    \centerline{\includegraphics[width=1\columnwidth]{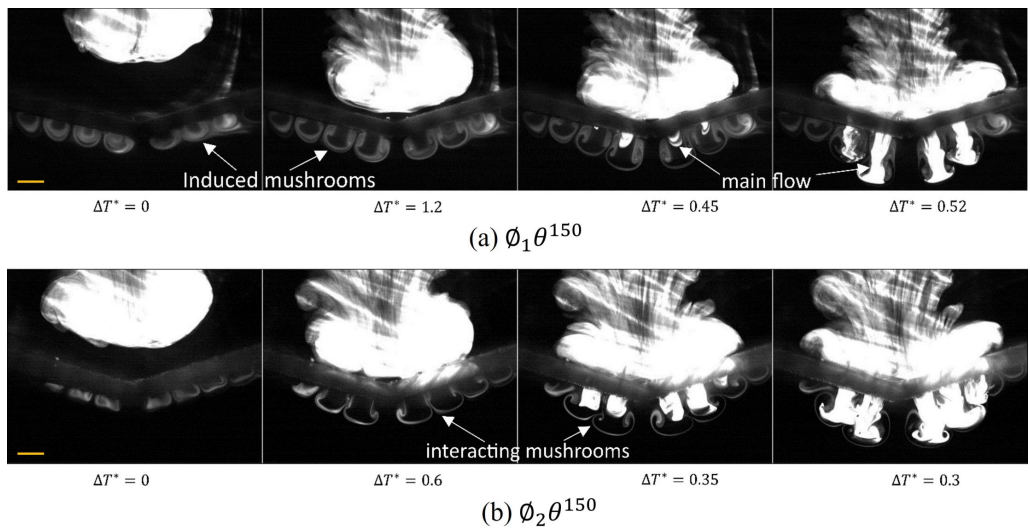}}
    \caption{The PLIF images for ${\phi_1\theta^{150}}$ and ${\phi_2\theta^{150}}$ depicting the initial jet instability due to mushroom formation. Here, $\Delta T^{*}$ is the difference in $T^*$ The scale represents 5 mm.}
    \label{figure:7}
\end{figure}

\begin{figure}
    \centering
    \centerline{\includegraphics[width=1\columnwidth]{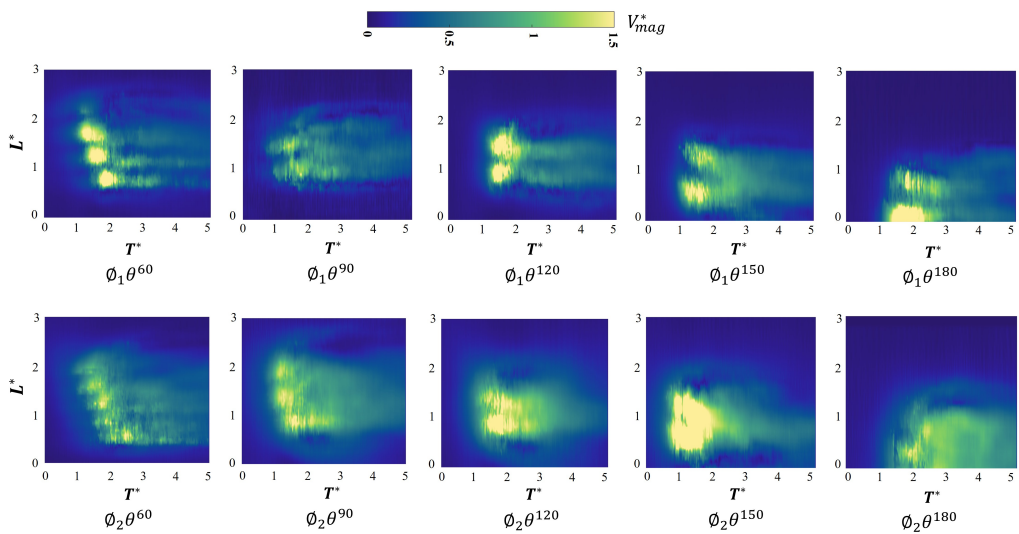}}
    \caption{The non-dimensional velocity magnitude (non-dimensionalised using free vortex ring velocity magnitude) contour plotted against ${T^{*}}$ and ${L^{*}}$ (= $L/D$ where \textit{L} is the length of the line about which the data is plotted) at a distance of 0.5\textit{D} from the perforated surface.}
    \label{figure:8}
\end{figure}

\begin{figure}
    \centering
    \centerline{\includegraphics[width=1\columnwidth]{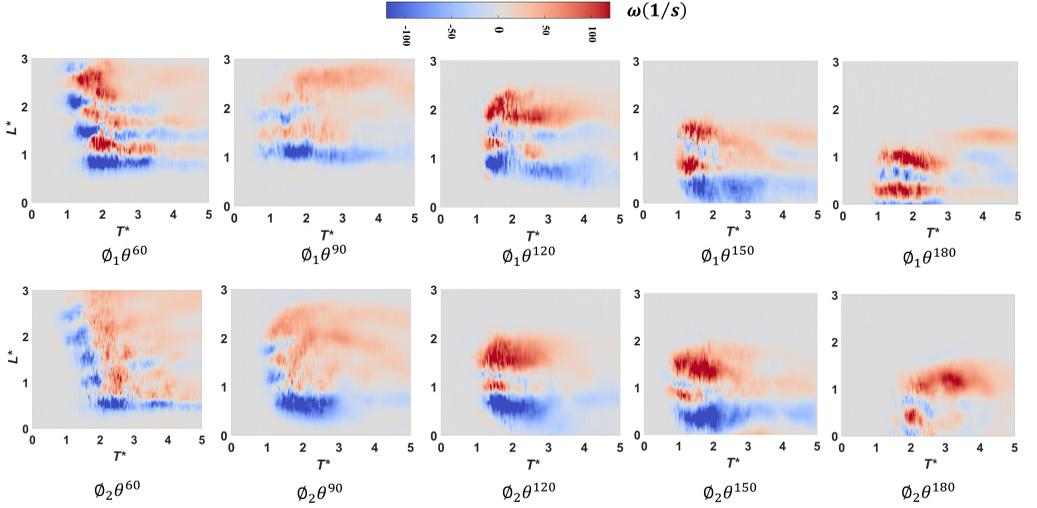}}
    \caption{The dimensional vorticity contour plotted against ${T^{*}}$ and ${L^{*}}$ (= $L/D$ where \textit{L} is the length of the line about which the data is plotted) at a distance of 0.5\textit{D} from the perforated surface.}
    \label{figure:9}
\end{figure}

For ${\theta}$ = 90\textdegree, three velocity peaks are seen followed by two peaks for the rest of the cases (figure 8). For cases with ${\phi_1}$, a distinct gap between the peaks can be observed that is not present in case of ${\phi_2}$. For ${\phi_2\theta^{150}}$, the peaks are seen to be merging with each other. Another important information that can be extracted from the plot is that for ${\theta}$ $\geq$ 120\textdegree\ , the first velocity peak occurs nearer to the apex and the second peak occurs away from it i.e., there is reversal in the sequence in which the jets come out (figure 8). This suggests that the vortex first starts to penetrate from the holes nearer to the apex of the perforated surface. Hence, for the present set of parameters, ${\theta}$ = 120\textdegree\ behaves like a transition point where the flow field changes and starts to tend towards $\theta$ = 180\textdegree. Moreover, with an increase in ${\theta}$, the overall flow map (figures 8 and 9) can be seen to be shifting towards $L^*$ = 0. For ${\phi_2}$, the vorticity zones are more circular in shape compared to ${\phi_1}$ where they are more linear. The overall vorticity dynamics will be discussed in the next section.  

\subsubsection{Vorticity dynamics}

\begin{figure}
    \centering
    \centerline{\includegraphics[width=1\columnwidth]{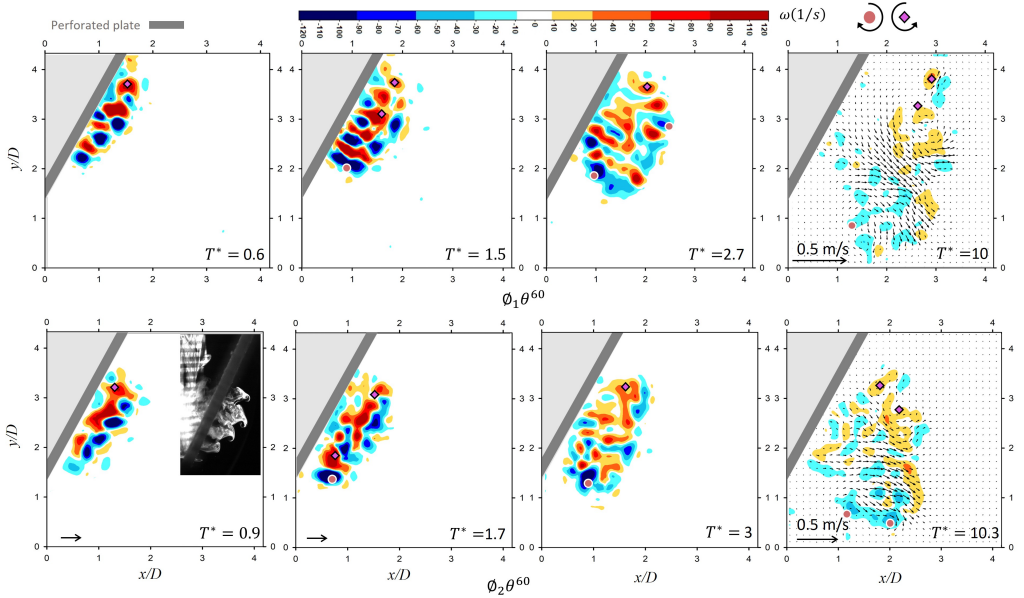}}
    \caption{Evolution of vorticity field for ${\phi_1\theta^{60}}$ and ${\phi_2\theta^{60}}$ plotted at different $T^*$ values. The scale vector indicates $0.5   m/s$. The pink dot represents the CCW rotating vortex cores and the green dots represent the CW rotating vortex cores detected using $\Gamma_2$ method.\textcolor{black}{The inset in the first plot of $\phi_2$ is the PLIF image of the interaction around that time.}}
    \label{figure:10}
\end{figure}

\begin{figure}
    \centering
    \centerline{\includegraphics[width=1\columnwidth]{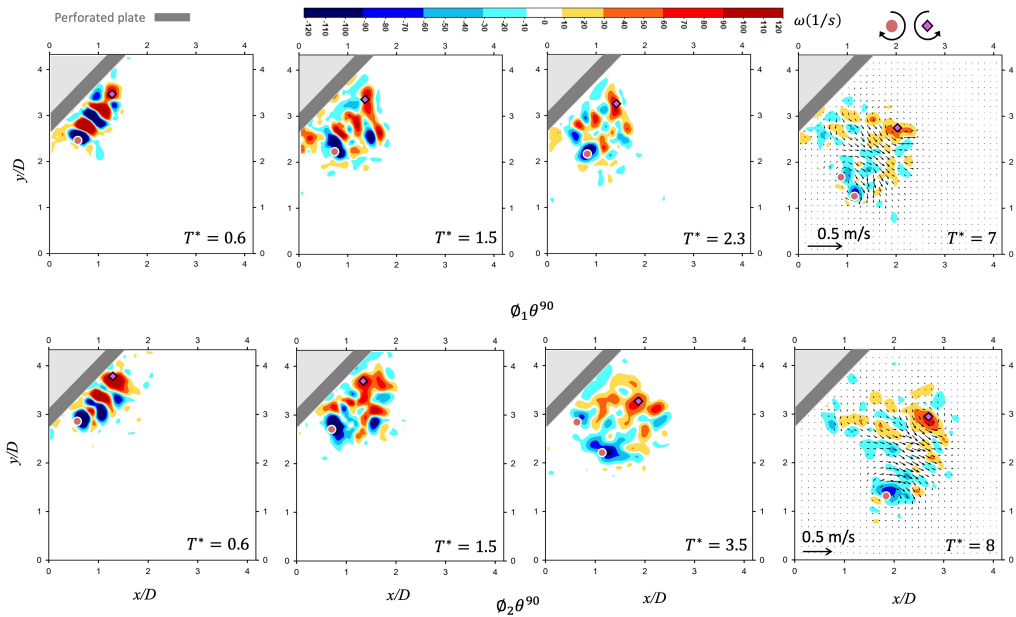}}
    \caption{Evolution of vorticity field for ${\phi_1\theta^{90}}$ and ${\phi_2\theta^{90}}$ plotted at different $T^*$ values. The scale vector indicates $0.5   m/s$. The pink dot represents the CCW rotating vortex cores and the green dots represent the CW rotating vortex cores detected using $\Gamma_2$ method.}
    \label{figure:11}
\end{figure}

We plot the vorticity contours obtained from the PIV data in figures 10-14 in increasing order of ${\theta}$. In all the final cases, the vortictiy contour is superimposed with the vector fields to give a sense of the flow. When a vortex ring interacts with a perforated plate of such kind, the emerging flow in the downstream remains highly chaotic. The reason for this being the conversion of the vortex ring to single jets and their further interaction with each other. For cases with ${\theta}$ $\leq$ 120\textdegree\ , we see this phenomenon occurring symmetrically on two halves of the surface whereas for ${\theta}$ $>$ 120 the jets on both sides start talking to each other. These features will be highlighted in this section where we discuss how the vortex cores move and interact in the downstream region. We denote the clockwise (CW) and counter (CCW) rotation of vortices by their centre using orange dots and pink rhombus respectively. The location of these symbols have been identified using the $\Gamma_2$ method as described in section 2.4. 

\begin{figure}
    \centering
    \centerline{\includegraphics[width=1\columnwidth]{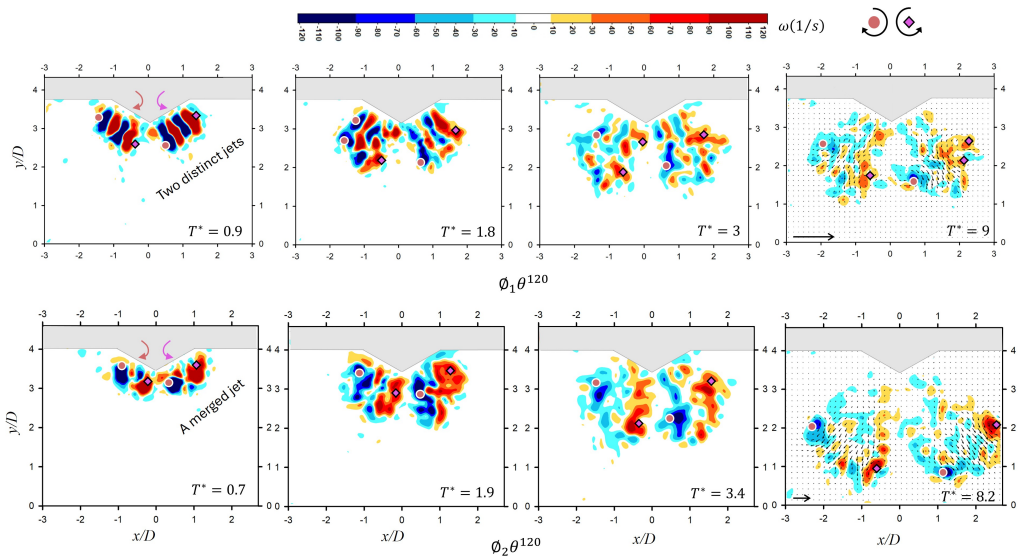}}
    \caption{Evolution of vorticity field for ${\phi_1\theta^{120}}$ and ${\phi_2\theta^{120}}$ plotted at different $T^*$ values. The scale vector indicates $0.5   m/s$. The pink dot represents the CCW rotating vortex cores and the green dots represent the CW rotating vortex cores detected using $\Gamma_2$ method.}
    \label{figure:12}
\end{figure}

For ${\phi_2\theta^{60}}$ we see a stretch of negative vorticity coming out of the surface compared to ${\phi_1\theta^{60}}$ where the features resemble more like broken individual jets (figure 10). A CCW vortex develops near the hole that is farthest to the apex. The vorticity remains more distributed in case of ${\phi_2\theta^{60}}$ and the flow moves at a slower speed. For ${\phi_1\theta^{60}}$ ultimately the flow converges to the development of two major vortical regions that propel in the direction nearly perpendicular to the perforated surface. For ${\phi_2\theta^{60}}$, the CW vortex that develops near the apex has more vorticity and propels faster than the CCW resulting in the horizontal convection of the flow. The faster motion of this CW vortex is a result of the final momentum that is imparted by the disintegration of the core near the apex. The two vortical regions at the extreme ends of the flow is seen to break up into multiple small-scale vortical zones. Similar is the case for ${\theta}$ = 90\textdegree (figure 11) where the jets are more organised with some rotating features and the rotating zones develop very early propelling in a nearly normal direction.
\begin{figure}
    \centering
    \centerline{\includegraphics[width=1\columnwidth]{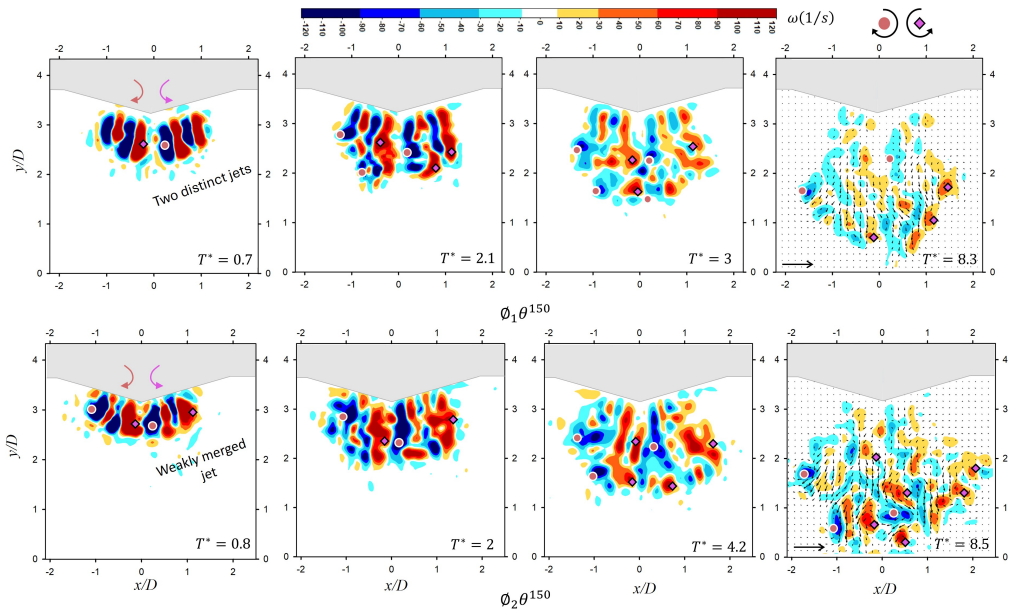}}
    \caption{Evolution of vorticity field for ${\phi_1\theta^{150}}$ and ${\phi_2\theta^{150}}$ plotted at different $T^*$ values. The scale vector indicates $0.5   m/s$. The pink dot represents the CCW rotating vortex cores and the green dots represent the CW rotating vortex cores detected using $\Gamma_2$ method.}
    \label{figure:13}
\end{figure}

We depict both sides of the flow for ${\theta}$ $\geq$ 120\textdegree to identify the $\theta$ at which the flow on the two halves starts to interact with each other. Two distinct jets with opposite vorticity at edges can be seen to develop initially around $T^{*}$ = 0.9 for ${\phi_1\theta^{120}}$ with the rotating regions on the outer and inner sides of the flow on each side (figure 12). For ${\phi_2\theta^{120}}$ instead of long jets, we see circular patches of vorticity developing initially around the same time resembling a merged jet. The out coming flow in ${\phi_2}$ already starts to interact due to the proximity of the holes resulting in larger circular patches. Interestingly, the vorticity cancellation process seems less intense in these larger vorticity zones (in ${\phi_2}$) compared to the jets (in ${\phi_1}$) resulting in the higher vorticity values of the core at a later instance. Due to a larger spreading area in case of ${\phi_2\theta^{120}}$, we can see at ${T^{*}}$ = 3.4, the inner vortical cores come very near to each other ultimately diverging in the flow direction.

\begin{figure}
    \centering
    \centerline{\includegraphics[width=1\columnwidth]{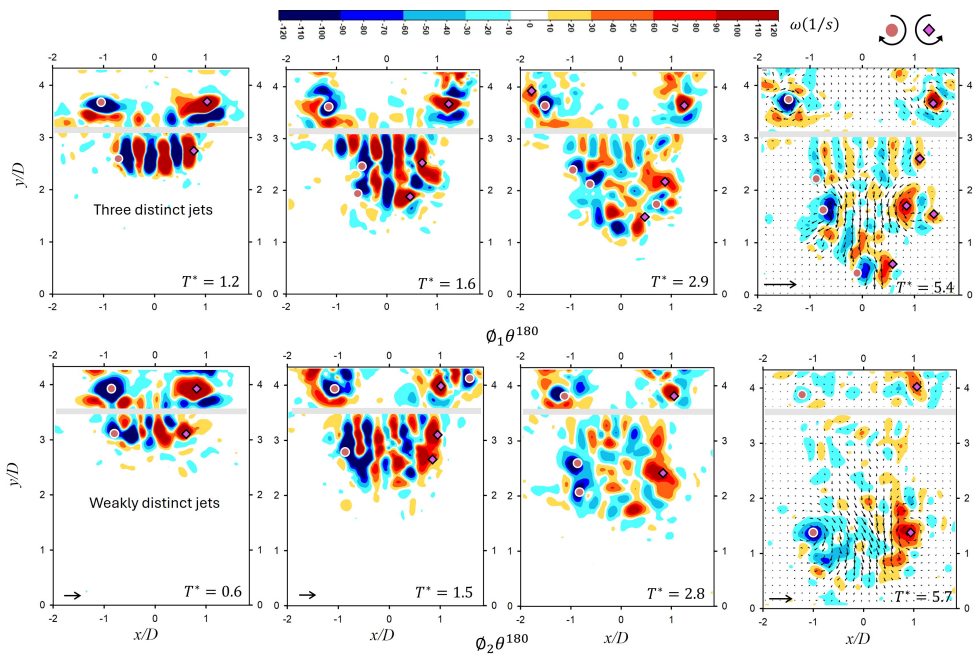}}
    \caption{Evolution of vorticity field for ${\phi_1\theta^{180}}$ and ${\phi_2\theta^{180}}$ plotted at different $T^*$ values. The scale vector indicates $0.5   m/s$. The pink dot represents the CCW rotating vortex cores and the green dots represent the CW rotating vortex cores detected using $\Gamma_2$ method.}
    \label{figure:14}
\end{figure}

For cases with ${\theta}$ = 150\textdegree (figure 13), the flow initially behaves similarly to the preceding case with much more proximity between the inner jets from other halves. Along with vorticity cancellation on each side of the surface, we see clear interaction of the vortex cores for ${\phi_1\theta^{150}}$ at ${T^{*}}$ = 3 and ${\phi_2\theta^{150}}$ at ${T^{*}}$ = 2.  Neat jets of fluid can be discerned for ${\phi_1\theta^{150}}$ that results in vorticity cancellation from adjacent jets and ultimately breaks into multiple vortical zones as shown in figure 13. However, a less coherent flow is observed for ${\phi_2\theta^{150}}$ resulting in the development of many vortical zones resembling turbulent flows. As shown in figure 6 (red marked zone), the vortex reformation in the far downstream does not happen in case of ${\theta}$ = 150\textdegree\ due to the interaction of flow from other halves.

For ${\theta}$ = 180\textdegree (figure 14) we show both the upstream and the downstream regions of the perforated surface. For other cases, the upstream flow was not depicted because of the uncertainty in the PIV data that can arise in cases with lower ${\theta}$ values due to wall effects. In this regard,  figure S5 (see supplementary file) is presented in the supplementary sheet showing the effect of $\phi$ and $\theta$ on the circulation of the vortex ring just before the interaction. On the upstream side, as the vortex approaches the surface boundary layer of opposite sense starts to develop that separates to form the secondary vortices of opposite sense \citep{naaktgeboren_krueger_lage_2012, PhysRevFluids.8.044701}. It can also be seen that the vortex core sustains for a longer period in case of $\phi_1$ which is expected due to lesser porosity. \textcolor{black}{For ${\phi_1\theta^{180}}$, the jets of opposite vorticity interact and shed a pair of vortex from the head in the central region (figure 14 ($T^{*}$ = 2.9 and 5.4)). This is due to the shearing between the central and the adjacent jets. This is also contributed by the mushroom shaped head which can be seen from PLIF images.} Along with these small structures, the overall flow preserves the axial bulk velocity at the central region resulting in the formation of two large vortical cores at the edges of the flow field \citep{PhysRevFluids.8.044701}. In contrast, we do not observe central vortices for  ${\phi_2\theta^{180}}$, due to early mixing of jets and two large-scale vortical structures form after the central vortices cancel each other. \textcolor{black}{It is important to mention that in our studies (present and \citep{PhysRevFluids.8.044701}) we use solenoid-generated single vortex ring with high $Re_\Gamma$ values and hence see very little fluid accumulation on the upstream region like \citet{xu_wang_feng_he_wang_2018}. The flow fields that evolve in the downstream region are highly sensitive to geometric parameters like $\phi$ and structure of the holes and, flow parameters like $Re_\Gamma$ of the vortex ring. Hence, a direct comparison cannot be carried out from the already existing works. However, features like jets and vorticity cancellation have been seen in almost all the studies at appropriate $Re_\Gamma$ and $\phi$ values. }

\begin{figure}
    \centering
    \centerline{\includegraphics[width=0.85\columnwidth]{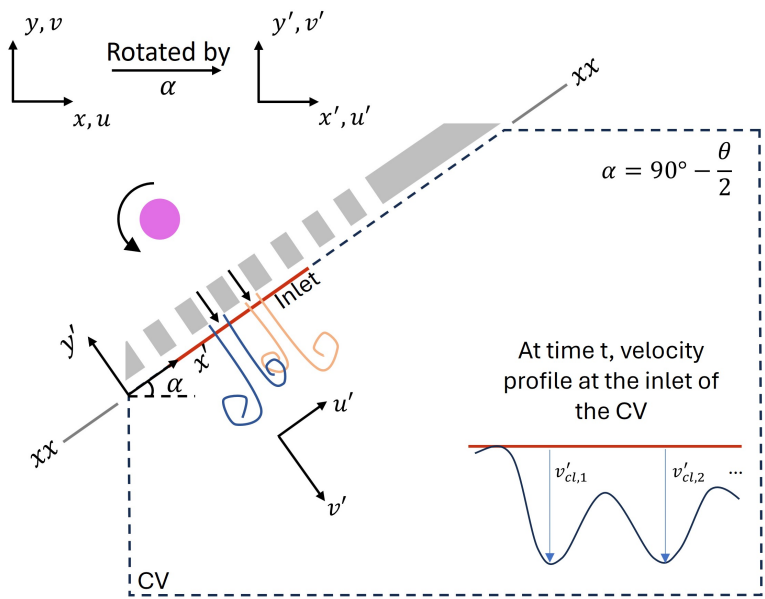}}
    \caption{\textcolor{black}{Schematic of the CV for the cumulative slug model. The global coordinate axis is rotated by an angle $\alpha$ to get a new coordinate axis in $x'y'$ coordinates. The figure shows the inlet of the CV marked with a red line and shows two jets (in blue and orange color) forming in the downstream region. A typical velocity profile about the inlet of the CV is depicted where $v'_{cl}$ represents the centre line velocity of each jet. The numbers in the subscript of the center line velocity depict the jet numbers.}}
    \label{figure:15}
\end{figure}
\subsubsection{Cumulative Slug flow model}

\textcolor{black}{The interaction of a vortex ring with perforated plate of appreciable thickness at high Reynolds number results in the formation of multiple jets and consequently a reformed vortex. Due to the constriction, the jets accelerate till a certain point beyond which they start decaying. We hypothesize these jets as impulsively started jets that contribute to the circulation growth in the downstream region. Figure 15 shows the control volume (CV) chosen for the validation of a cumulative 1D slug model with the global coordinates axis (x,y) rotated by angle $\alpha$ to form a new coordinate system $(x',y')$. According to the standard 1D slug model \citep{Didden1979} the circulation is produced by the transport of vorticity across the inlet plane (marked red in Figure 15) for $y'$$<$0 inside the enclosed CV. The vorticity flux is given as}

\textcolor{black}{\begin{equation}
    \frac{d\Gamma}{dt} =\int_{0}^{\infty}\omega'v'dx'=\int_{0}^{\infty}(\frac{dv'}{dx'}-\frac{du'}{dy'})v'dx'
\end{equation}}

\textcolor{black}{Invoking the parallel jet assumption and neglecting the second term that comes due to the gradient of radial velocity in the axial direction, the vorticity flux for each jet can be written as:}

\textcolor{black}{\begin{equation}
\frac{d\Gamma}{dt} = \int_{0}^{\infty}v'\frac{dv'}{dx'}dx' = \frac{1}{2}v'^2(t)
\end{equation}}

\textcolor{black}{Equation (3.2) is the standard slug model that is used to estimate the evolution of circulation in time where $v'$ is considered as the center line velocity of the jet \citep{Rosenfeld_1998}. In our problem, since we have multiple jets coming out of the holes, the total circulation growth in the downstream region should be a result of the circulation produced by the individual jets. Thus, we propose a cumulative slug model where the overall time evolution of circulation is estimated by the sum of the circulation of each of the jets. Then by considering the centre line velocity or the peak velocity of each jet, the formulation for the time evolution of cumulative circulation is given as}

\textcolor{black}{\begin{equation}
    \frac{d\Gamma_c}{dt} =\frac{1}{2}\sum_i(v'_{cl})_i^2(t)
\end{equation}}

\textcolor{black}{where $\Gamma_c$ is the cumulative circulation,  $(v'_{cl})_i$ is the centre line velocity of each individual jet \textit{i} . Both the left and the right hand side of equation (3.3) has been evaluated using the obtained PIV data and has been plotted in figure 17. As shown in figure 16, the velocity profile about the inlet of the CV varies with time and exhibit different number of peaks for different $\theta$ and $\phi$ values. For $\phi_1$, the jets are sufficiently apart to get multiple distinct peaks of velocity profile however, for $\phi_2$, the pitch of the perforated plate being small, the jets do not exhibit distinct peaks for $\theta$ = 120\textdegree\ and 150\textdegree. For $\theta$ = 180\textdegree\ we see three sharp peaks very near to each other. We validate the slug model only for $\theta$ $\geq$ 120\textdegree\ since for $\theta$ lower than that, the assumption of parallel jets could not be justified. For such flow condition, due to the steepness of the plates, the flow doesn't come out orthogonal to the plate.}

\textcolor{black}{A quantitative comparison of the non-dimensional circulation generated by the positive and negative vorticity in the downstream region is presented in figure 17 for each of the cases. The evolution of circulation in time obtained from equation (3.3) is also plotted in the same plots for $\theta$ $\geq$ 120\textdegree. Owing to symmetry, we consider the right side of the apex axis for all the cases (as shown in figure 1(c)). The pink line depicts the circulation generated by the positive vorticity ($\Gamma^+$, CCW) and the green line shows the circulation generated by the negative vorticity ($\Gamma^-$,CW). Since the right-hand side of the flow is considered, the side of the vortex ring having CCW sense about the $xy$ plane is discussed.}

\textcolor{black}{Intuitively, the fluid emerging out should have a natural tendency to generate higher positive circulation in CCW direction. However, this is not the case for all the $\theta$. For $\theta$$\leq$ 90\textdegree, the positive circulation remains marginally above the negative circulation. Whereas for $\theta$ = 120\textdegree\ and 180\textdegree\ the positive and negative circulation lines almost overlap each other. This indicates that the flow after passing through the holes reconfigure itself and generate both the vorticity. The effect of placing an inclined perforated plate is minimal from the point of view of a obtaining a symmetric circulation field in the downstream region and hence a symmetric reformed vortex ring. For $\theta$ = 150\textdegree\ as seen in previous figures (4, 5 and 13), the flow from each side begins to interact after emerging out resulting in vorticity cancellation at the central region in addition to the cancellation between the adjacent jets on each sides. In figure 17, it can be seen that the pink and green line diverges after a point of time for $\theta$ = 150\textdegree. For $\phi_1$ this divergence takes place much after the global decay of the circulation begins whereas for $\phi_2$ this coincides with the global decay of the circulation.}

\textcolor{black}{For $\theta$ $\geq$ 120\textdegree, the cumulative slug model overlaps both the $\Gamma^+$ and $\Gamma^+$ lines that depicts the circulation growth in the downstream region. It can be assumed that the jets preserve their flow propoerties in the near field region \citep{Rui_ni_2023_jets}. The jets do not start to interact till a certain period of time and hence can be assumed to be parallel jets with individual jets contributing to the overall growth. Beyond the peak, due to the self interactions and viscous dissipation the decay begins resulting in decrease in both the circulation values. The slug model doesn't work beyond this peak since it does not take into account the dissipation.}

\begin{figure}
    \centering
    \includegraphics[width=1\linewidth]{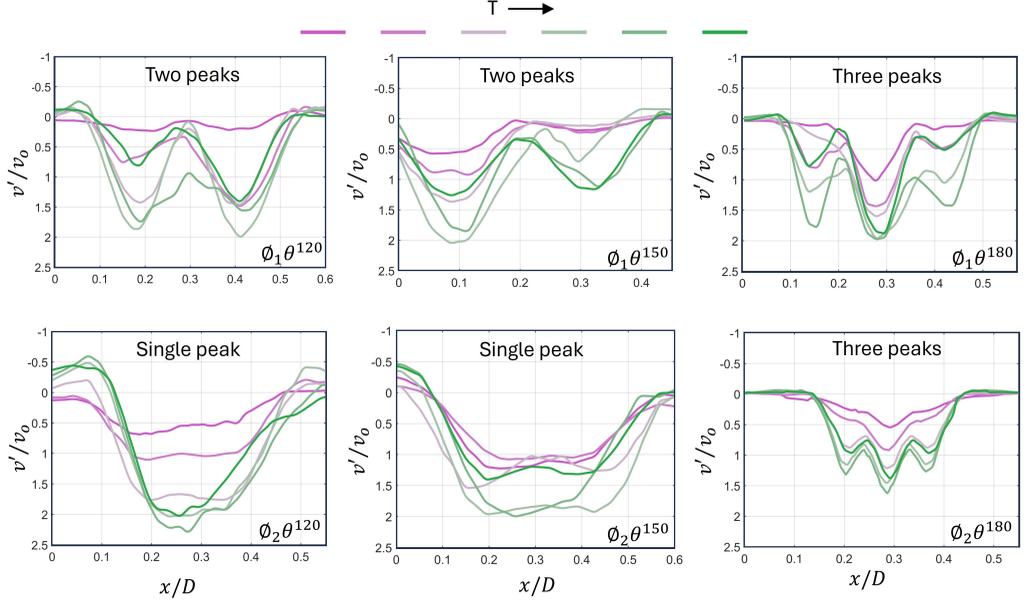}
    \caption{\textcolor{black}{The velocity ($v'$) profile about the inlet of the CV at different time duration normalised by the convective velocity of the free vortex ring. These profiles are shown only during the period when the cumulative slug model is valid. The number of peaks considered for each case in the cumulative slug model is mentioned within respective plots.}}
    \label{fig:enter-label}
\end{figure}

\begin{figure}
    \centering
    \centerline{\includegraphics[width=1\columnwidth]{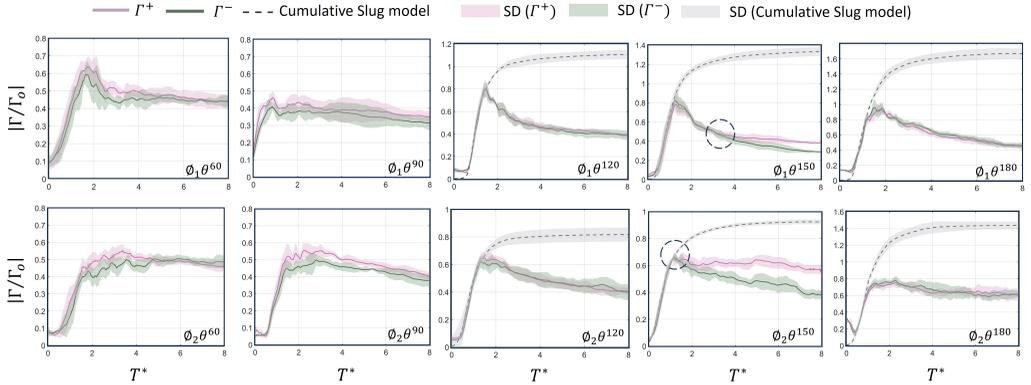}}
    \caption{\textcolor{black}{The variation of absolute values of non-dimensional positive ($\Gamma^+$) and negative $\Gamma^-$) circulation (non-dimensionalised by $\Gamma_0$ for each case) in the downstream region of the flow field after interaction for all the cases with $T^{*}$. For $\theta$$\geq$120\textdegree, the cumulative slug model is plotted using equation (3.3). The circular insets highlight the instance where the $\Gamma^+$ and $\Gamma^-$ value starts to diverge. Here, $T^{*}$ = 0 represents the instance at which the velocity vectors become detectable after interaction in the downstream region. SD represents the standard deviation of the data.}}
    \label{figure:17}
\end{figure}

\begin{figure}
    \centering
    \centerline{\includegraphics[width=1\columnwidth]{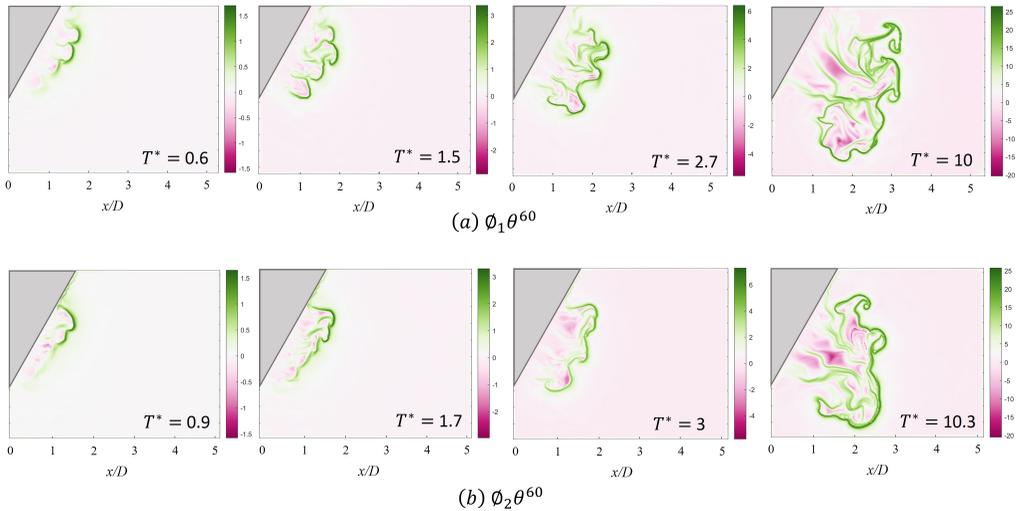}}
    \caption{Backward FTLE field for (a) ${\phi_1\theta^{60}}$ and (b) ${\phi_2\theta^{60}}$ shown at different $T^{*}$.}
    \label{figure:19}
\end{figure}

\subsubsection{Lagrangian analysis}

Figure 18, 19, 20 and S6 shows the backward FTLE fields for $\theta$ = 60\textdegree\ , 120\textdegree, 150\textdegree, and 180\textdegree\ respectively. Our primary aim is to elucidate if the jets emerging out of the permeable surface mix at all and if it does, how and when. Although a lot has been discussed in detail about the interaction of jets in the above sections, we finally want to conclude our analysis with a qualitative evaluation of the FTLE fields digging deeper into the dynamics of out coming jets from the holes. The LCS are the ridges of the FTLE fields that are transverse to the direction of minimum curvature \citep{Haller2001}. Based on the system's dynamical behavior, the LCS in the flow fields represent surfaces that does not allow transport across it. We plot the FTLE values by integrating backward in time that yields attractive FTLE ridges. Since we use different integration times for each of the $T^{*}$, the FTLE range obtained for each of the frames varies. However, bringing out the green ridges to understand the mixing dynamics remains the central theme of this exercise.
 \begin{figure}
    \centering
    \centerline{\includegraphics[width=1\columnwidth]{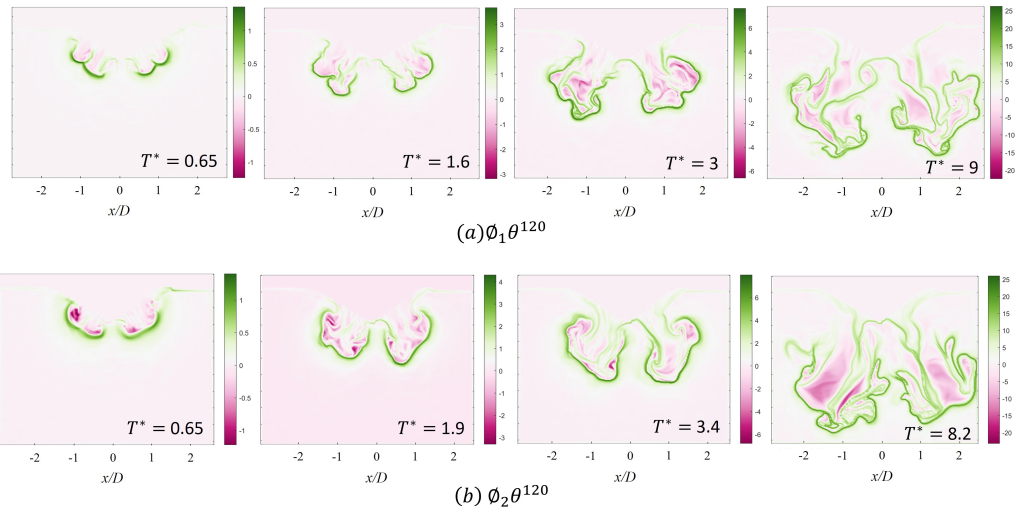}}
    \caption{Backward FTLE field for (a) ${\phi_1\theta^{120}}$ and (b) ${\phi_2\theta^{120}}$ shown at different $T^{*}$.}
    \label{figure:20}
\end{figure}

\begin{figure}
    \centering
    \centerline{\includegraphics[width=1\columnwidth]{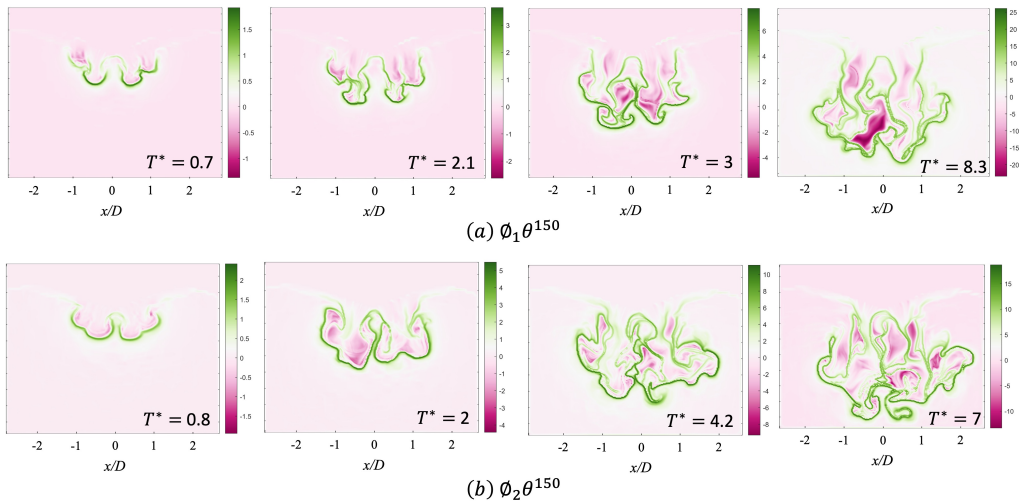}}
    \caption{Backward FTLE field for (a) ${\phi_1\theta^{150}}$ and (b) ${\phi_2\theta^{150}}$ shown at different $T^{*}$.}
    \label{figure:21}
\end{figure}

 From the first instance, green ridges between the jets coming out of the perforated surface can be detected in case of ${\phi_1\theta^{60}}$ unlike that for ${\phi_2\theta^{60}}$ (figure 18) where a continuous front of the attractive LCS develops. The ridge front of the flow structure is darkest (green) at the farthest side of the apex due to the early emergence of flow for ${\theta}$ = 60\textdegree\ as discussed earlier. For ${\phi_1\theta^{60}}$, the separation between the jets can be clearly observed at $T^{*}$ = 1.5 which further starts to blur out with the passage of time. However, for ${\phi_2\theta^{60}}$ we see a very less amount of green separatrices appearing between the jets indicating towards the mixing of flow at very early stages of emerging out. The flow structures at the later stages correspond well with the vorticity contours in figure 10. Similar patterns can be observed for ${\theta}$ = 120\textdegree (figure 19) with highly mixed flows for ${\phi_2\theta^{120}}$. We see no interaction between the ridges of the other halves of the surface and neat vortical structures develop. In case of ${\phi_1\theta^{150}}$  (figure 20), the interaction of the flow from the other halves can be seen around $T^{*}$ = 3 whereas this begins much early for ${\phi_2\theta^{150}}$. At further instances, very chaotic flow with a random distribution of attractive ridges develops which was educed from PLIF and PIV results. \textcolor{black}{The FTLE fields for $\theta$ = 180\textdegree\ have also been reported by \citet{xu_wang_feng_he_wang_2018} however, the $Re_\Gamma$ value of the vortex ring in their study was relatively smaller resulting in partial penetration of the vortex core and organized FTLE structures. Since the value of $\theta$ was fixed to 180\textdegree\ and under laminar conditions, no discussion on the mixing of flow was reported.} Finally, with the attractive LCS we could successfully explain the various events taking place during the interaction process. Although we do not show the forward FTLE or the repelling ridges, it can effectively be utilized to understand the entrainment quality associated with the flow structures that reform in the downstream region after interacting with the perforated surface. Furthermore, a rigorous quantitative analysis of the ridges can bring out more insights into the mixing phenomenon among the jets.

\section{Summary and Conclusions}
 
Experiments were conducted to explore the interaction dynamics of a vortex ring ($Re_{\Gamma}$ = 10500) impinging on perforated surface (${\phi_1}$ = 0.24 and ${\phi_2}$ = 0.44) with different included angles (${\theta}$ = 60\textdegree - 180\textdegree) using PLIF, PIV, Lagrangian FTLE and $\Gamma_2$ vortex identification techniques. By changing the the value of $\theta$ the interaction process is altered significantly. The early interaction of the vortex ring with the perforated surface results in a sequential ejection of the fluid and a reduction in the number of ejecting holes with the increase in the ${\theta}$ value. Due to higher proximity of the holes in case of ${\phi_2}$, the jets are seen to interact immediately as they start to emerge out of the perforated surface. Whereas for ${\phi_1}$, the jets interact much later and their development is characterized by the occurrence of K-H instability at its edges. 

\textcolor{black}{An intriguing phenomenon has been observed involving the formation of mushroom-shaped structures, prompting the question of whether vortex rings should be considered as discrete systems i.e.e without their induced flow fields, as has been the approach in interaction studies. The fluid ahead of the vortex ring moving with the induced velocity crosses the perforated surface earlier than the main vortex resulting in the growth of dormant mushroom structures which is later caught by the faster main flow. These structures are a result of placing a perforated plate in the path of a vortex ring and impart initial instability to the jet structures.}

\textcolor{black}{Another interesting finding includes the overlapping of the positive and negative circulation values in the downstream region for one side of the perforated plates for $\theta$ $\leq$ 120\textdegree. This shows that splitting the vortex ring in two halves through perforated plate can reconfigure the flow with no bias of the upstream vorticity. Further, flow from each side was seen to interact for $\theta$ = 150\textdegree\ that results in additional vorticity annihilation at the central region along with those between adjacent jets. Furthermore, we propose a cumulative slug model using PIV data which takes into account the superposition of circulation growth by individual jets. We validate this cumulative model for $\theta$ $\geq$ 120\textdegree. Given the inlet velocity profiles, the growth of circulation can be predicted using this cumulative slug flow model. The Lagrangian FTLE fields finally confirmed towards the sequential emerging, early interaction of flow from holes in case of $\phi_2$ and interaction of flow from other halves resulting in turbulent structures with random attractive LCS.}

\textcolor{black}{Perforated plates play a crucial role in various industries, particularly in applications such as industrial processes,  flow modeling and drag reduction. However, real-world conditions are seldom ideal; flows are often turbulent, and perforated plates can vary in shape and porosity. Therefore, this study is significant for understanding the complexities of flow interactions in these non-ideal situations.} By using a perforated plate with different $\theta$ values, a reformed vortex can be generated from each side i.e., a single vortex ring can generate two reformed vortex rings. \textcolor{black}{Likewise, n vortex rings can be generated by interacting a single vortex ring with n-faced perforated surfaces.} Further, this generation of reformed vortex rings can also be suppressed by maintaining the $\theta$ value near 150\textdegree\ between the surfaces. Such results can prove to be useful for flow control and manipulation purposes.

\vspace{20pt}

\textbf{Acknowledgement}
\vspace{8pt}

S. Jain and S.J. Rao would like to thank the Prime Minister Research Fellowship (PMRF) for the financial support.
\vspace{8pt}

Declaration of Interests: The authors report no conflict of interest.

\bibliographystyle{jfm}
\bibliography{jfm-instructions}

\begin{figure}
    \centering
    \centerline{\includegraphics[width=1\columnwidth]{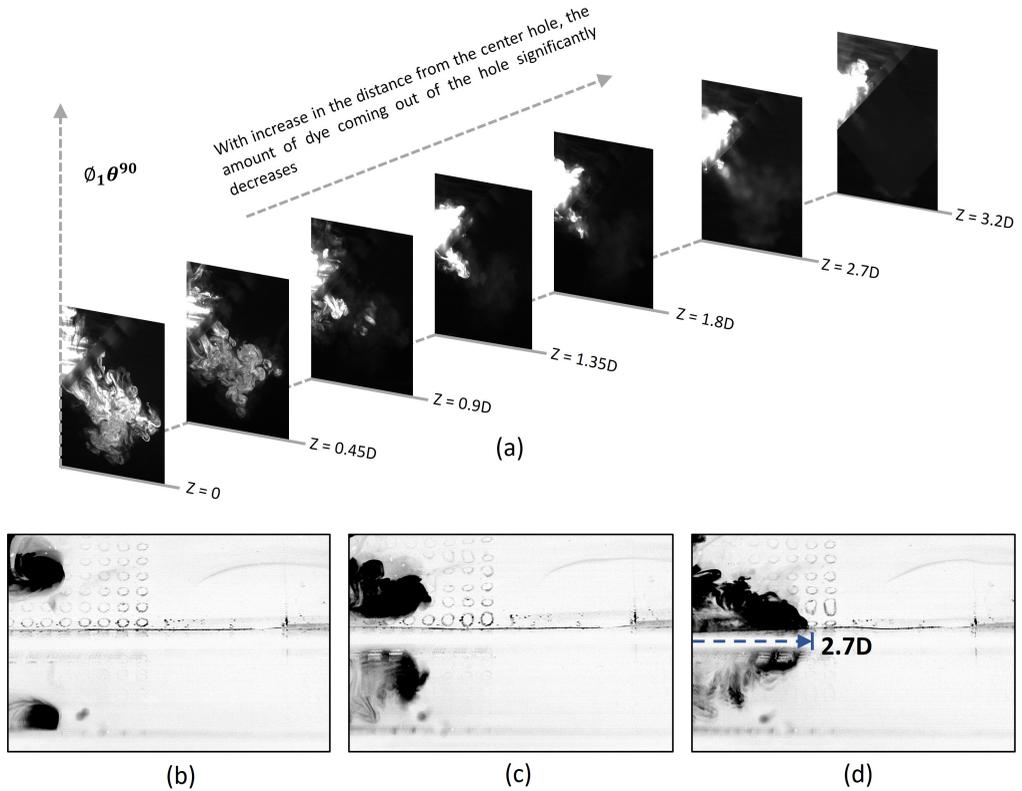}}
    \caption{\textbf{Figure S1} (a) PLIF images after interaction of vortex ring with perforated surface for ${\phi_1\theta^{90}}$ captured at different planes in z direction measured from the hole aligned with the vortex ring’s translational axis. For 1.35D and beyond, the images are captured at later times since the fluid comes out of the farther holes much later after the interaction at 0D takes place. (b), (c) and (d) are captured from an angle of around 45° from the vertical axis to capture the flow qualitatively in z direction. At 2.7D (see (a)), a very small chunk (the brightest part that is in focus) of fluid can be seen to come out. For the last hole, almost no flow comes out of the holes. From (a) and (d) it can be inferred that when the flow reaches the plate end, it almost ceases to come out of the holes.}
    \label{figure:S1}
\end{figure}

\begin{figure}
    \centering
    \centerline{\includegraphics[width=1\columnwidth]{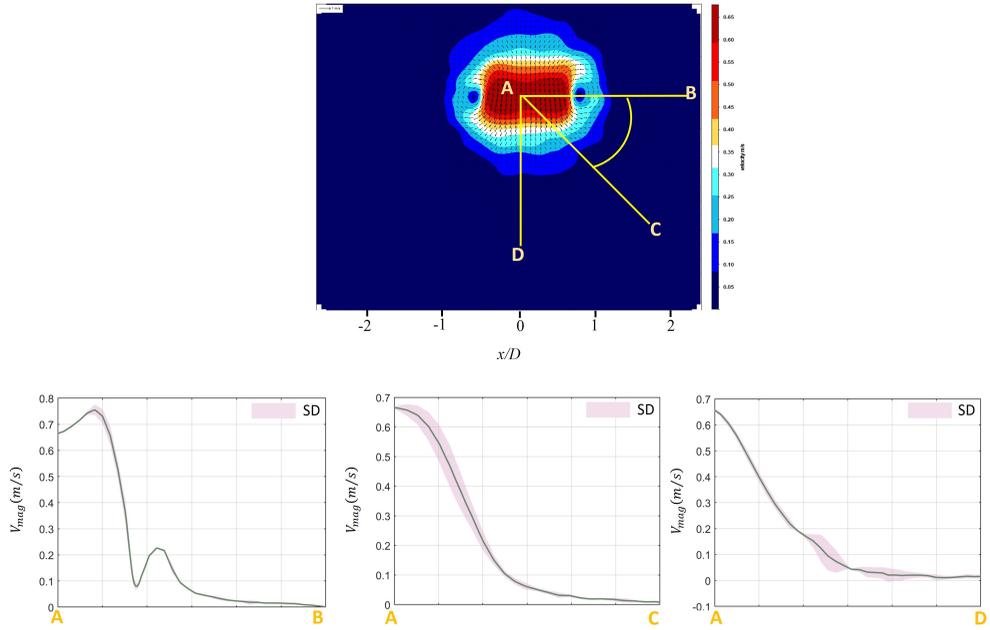}}
    \caption{\textbf{Figure S2} The velocity magnitude (${V_{mag}}$ along the lines AB, AC and AD where A corresponds to the center of the vortex ring. The angle between AB-AC and AC-AD is 45\textdegree. The shaded region represents the standard deviation (SD).}
    \label{figure:S2}
\end{figure}

\begin{figure}
    \centering
    \centerline{\includegraphics[width=0.75\columnwidth]{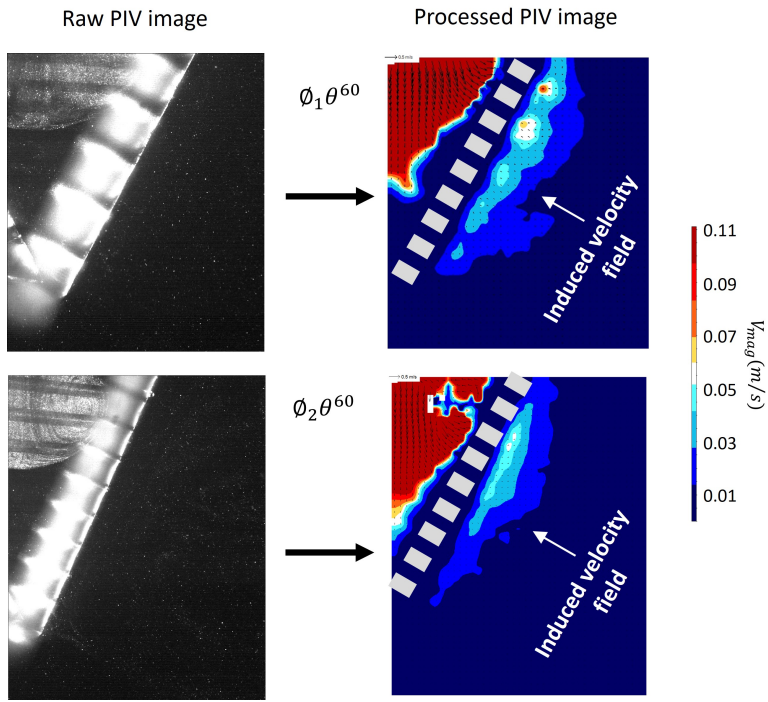}}
    \caption{\textbf{Figure S3} The induced velocity field across the perforated surface for $\theta$=60\textdegree is shown. The left column represents the raw PIV images, and the corresponding processed image is shown in the right column.}
    \label{figure:S3}
\end{figure}

\begin{figure}
    \centering
    \centerline{\includegraphics[width=1\columnwidth]{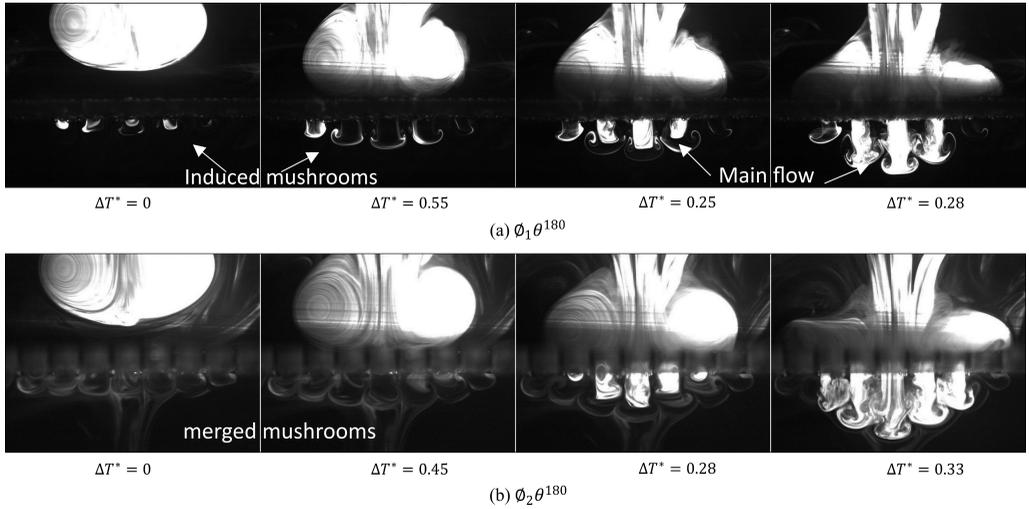}}
    \caption{\textbf{Figure S4} The PLIF images for ${\phi_1\theta^{180}}$ and ${\phi_2\theta^{180}}$ depicting the initial jet instability due to
mushroom formation. Here, $\Delta^*$ is the difference in $T^*$. The scale represents 5 mm.}
    \label{figure:S4}
\end{figure}

\begin{figure}
    \centering
    \centerline{\includegraphics[width=0.7\columnwidth]{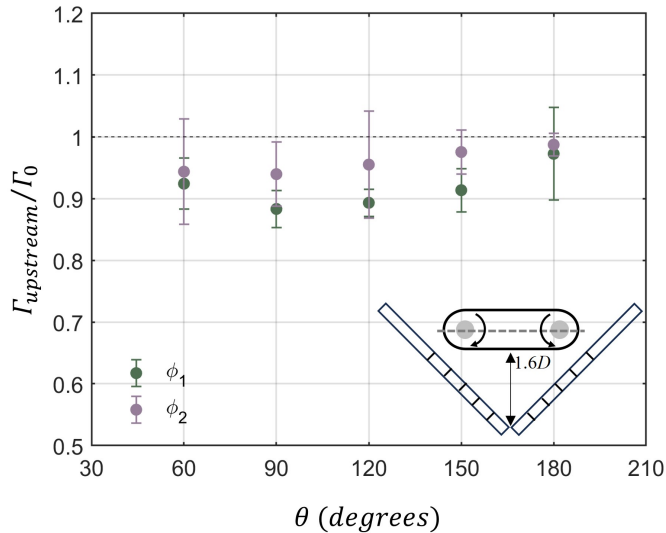}}
    \caption{\textbf{Figure S5} : The circulation of the vortex ring just before the interaction as shown in the inset normalized with the circulation of free vortex ($\Gamma_0$). The error bars represent the standard deviation (SD). For all the case, a minimal change in the circulation value is observed with respect to $\Gamma_0$ with a maximum reduction of around ~ 10\%. The values are particularly lower for the case of $\phi_1$ at all the $\theta$ which has a lower open area and behaves more like impermeable wall compared to $\phi_2$. It is expected that larger open area cases will behave more like free vortex case (i.e., without any surface) and the values should be nearer to 1. As the $\theta$ value reaches 180\textdegree, the ratio approaches the value 1 for both the case. The presence of the wall near the vortex ring before the interaction as shown in the inset contaminates the PIV data and hence, we see higher values of SD values.}
    \label{figure:S5}
\end{figure}

\begin{figure}
    \centering
    \centerline{\includegraphics[width=1\columnwidth]{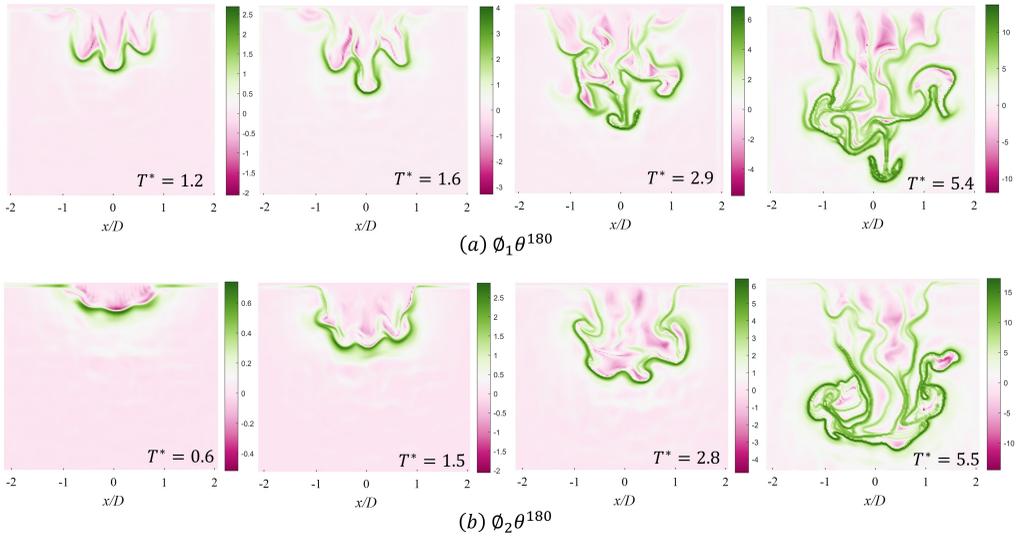}}
    \caption{\textbf{Figure S6}  Backward FTLE field for (a) ${\phi_1\theta^{180}}$ and (b) ${\phi_2\theta^{180}}$ shown at different $T^*$ corresponding to figure 14 (in the main manuscript).
}
    \label{figure:S6}
\end{figure}

\end{document}